\documentclass[journal=jacsat,manuscript=article]{achemso}

\usepackage{graphicx}
\usepackage[space]{grffile}
\usepackage{latexsym}
\usepackage{textcomp}
\usepackage{longtable}
\usepackage{multirow,booktabs}
\usepackage{amsfonts,amsmath,amssymb}
\usepackage{natbib}
\usepackage{url}
\usepackage{hyperref}
\hypersetup{colorlinks=false,pdfborder={0 0 0}}
\newif\iflatexml\latexmlfalse

\usepackage[english]{babel}

\usepackage[version=3]{mhchem} % Formula subscripts using \ce{}

\iflatexml
\documentclass[10pt,a4paper]{article}
\fi
\usepackage[latin1]{inputenc}
\usepackage{amsmath}
\usepackage{amsfonts}
\usepackage{amssymb}
\usepackage{graphicx}
\usepackage{multirow}

\author{Sung-Jin Kim}
\author{Adri{\'a}n Jinich}
\author{Al{\'a}n Aspuru-Guzik}
\affiliation{Department of Chemistry and Chemical Biology, Harvard University}
\email{aspuru@chemistry.harvard.edu}

\title[]{MultiDK: A  Multiple Descriptor Multiple Kernel Approach for Molecular Discovery and Its Application to The Discovery of Organic Flow Battery Electrolytes}

\begin{document}
%%%%%%%%%%%%%%%%%%%%%%%%%%%%%%%%%%%%%%%%%%%%%%%%%%%%%%%%%%%%%%%%%%%%%
%% The manuscript does not need to include \maketitle, which is
%% executed automatically.  The document should begin with an
%% abstract, if appropriate.  If one is given and should not be, the
%% contents will be gobbled.
%%%%%%%%%%%%%%%%%%%%%%%%%%%%%%%%%%%%%%%%%%%%%%%%%%%%%%%%%%%%%%%%%%%%%

\begin{abstract}
We propose a multiple descriptor multiple kernel (MultiDK) method for efficient molecular discovery using machine learning. We show that the MultiDK method improves both the speed and the accuracy of molecular property prediction. We apply the method to the discovery of electrolyte molecules for aqueous redox flow batteries. Using \emph{multiple-type - as opposed to single-type - descriptors}, more relevant features for machine learning can be obtained. Following the principle of the 'wisdom of the crowds', the combination of multiple-type descriptors significantly boosts prediction performance. Moreover, MultiDK can exploit irregularities between molecular structure and property relations better than the linear regression method by employing multiple kernels - more than one kernel functions for a set of the input descriptors. The multiple kernels consist of the Tanimoto similarity function and a linear kernel for a set of binary descriptors and a set of non-binary descriptors, respectively. Using MultiDK, we achieve average performance of $r^2 = 0.92$ with a set of molecules for solubility prediction. We also extend MultiDK to predict pH-dependent solubility and apply it to solubility estimation of quinone molecules with ionizable functional groups as strong candidates of flow battery electrolytes. 
\end{abstract} 

\section{Introduction}
Aqueous organic flow batteries are emerging as a low-cost alternative to store renewable energy \cite{huskinson_metal-free_2014, yang_inexpensive_2014, lin_alkaline_2015, AENM_201501449, winsberg_polytempo_zinc_2016}. For example, Huskinson et al., Yang et al., and Liu et al. experimentally showed that high capacity energy storage can be achieved using earth abundant organic electrolytes such as quinone molecules \cite{soloveichik_2015, yang_high-performance_2016}. Given the vast molecular space covered by all possible quinone molecules, high-throughput computational screening \cite{pyzer-knapp_bayesian_2016,plessow_trends_2016,peplow_materials_2015,santos_screened_2015,ma_deep_2015,shu_simulated_2015,hachmann_lead_2014,curtarolo_high-throughput_2013,kanal_efficient_2013,sokolov_computational_2011,fischer_predicting_2006,shoichet_virtual_2004,bajorath_integration_2002} is important to find electrolytes that satisfy the stringent requirement of aqueous flow batteries. In particular, the flow battery system in \cite{huskinson_metal-free_2014} requires a redox potential greater than 0.9V for a catholyte and less than 0.2V for an anolyte, as well as a solubility greater than one molar for both electrolytes. Moreover, quinone electrolytes operating in acid (pH 0) and alkaline (pH 14) flow battery environments were demonstrated in \cite{huskinson_metal-free_2014} and \cite{lin_alkaline_2015}, respectively. 

Recent high-throughput computational screening of benzo-, naphtho-, anthra-, and thiopheno-quinone libraries \cite{er_computational_2015, pineda_flores_bio-inspired_2015} demonstrated that the reduction potential of these redox couples can be predicted accurately utilizing molecular quantum chemistry methods and linear regressions. Using the free energy of solvation as a proxy descriptor, the molecular solubility of electrolytes was also predicted in both references. Here, we build upon this work by developing a machine learning strategy that results in strong correlations with experimental solubility data predicts the required molecular properties in order to accelerate molecular screening by several orders of magnitude.

The computational prediction of molecular solubility has been a research topic for decades, with most research being driven by the field of drug discovery \cite{wang_recent_2011,skyner_review_2015,soloveichik_2015}. However, predicting the solubility of organic electrolytes is particularly challenging, given the stringent target solubilities and the extreme pH values of flow battery electrolyte solutions \cite{huuskonen_estimation_2000}. While the target solubility of drug molecules is generally less than 0.1 molar, the target for flow battery organic electrolytes can be more than 1 molar. Moreover, molecular libraries to screen potential flow battery electrolytes include extremely acidic \cite{huuskonen_estimation_2000} or basic organic molecules \cite{lin_alkaline_2015} while the majority of drug candidates are relatively weak acids and bases \cite{wang_recent_2011,bhal_rule_2007,fischer_predicting_2006,bergstrom_accuracy_2004}. 

Both machine learning and quantum chemical approaches can be used to estimate molecular solubility. Whereas machine learning approaches predict solubility based on training to experimental data \cite{mitchell_machine_2014, hughes_why_2008, palmer_random_2007}, quantum chemistry aims to predict solubility from first principles \cite{er_computational_2015, mcdonagh_uniting_2014, marten_new_1996, tannor_accurate_1994}. Although quantum chemical approaches are preferable for obtaining a mechanistic understanding of underlying principles \cite{mcdonagh_uniting_2014, skyner_review_2015}, our focus here is on machine learning approaches which facilitate high-throughput and artificially-intelligent molecular discovery \cite{mitchell_machine_2014, raccuglia_machine-learning-assisted_2016, silver_mastering_2016}.

Machine learning approaches can be categorized into three types of methods according to the types of descriptors used: property-based methods, structure-based methods, and functional group-based methods (Table \ref{table:sol_chart}). Property-based methods predict physicochemical values based on molecular properties which can be measured experimentally or obtained from computational approaches. One such property used for solubility estimation is the partition coefficient, the logarithm of which is denoted as logP \cite{jain_estimation_2001, ran_estimation_2002, delaney_esol:_2004, Wang_etal_2009}. Several methods have been proposed to calculate logP \cite{tetko_application_2004, tetko_prediction_2001, lipinski_experimental_2001, viswanadhan_atomic_1989}. The general solubility estimation method (GSE), with its extended and modified variants, is an example of a property-type method which estimates logS from logP \cite{jain_estimation_2001, ran_estimation_2002, delaney_esol:_2004, Wang_etal_2009, ali_revisiting_2012}. On the other hand, structure-based methods rely on the estimation of solubility as a function of molecular structure. Structure is usually represented by a binary fingerprint, consisting of molecular topology, connectivity, or fragment information \cite{zhou_scores_2008, durant_reoptimization_2002}. Finally, group-based methods partition molecules into functional groups, and the contribution of each to the value of a physicochemical property is estimated \cite{klopman_estimation_1992, kuhne_group_1995, cheng_binary_2011}.  

Property-based methods generally involve fewer regression parameters than the other two approaches, but require additional computation in order to estimate intermediate properties included in the descriptor set. If large experimental data is available for intermediate properties such as logP, property-based methods can predict solubility for a wider range of molecules than any of the other methods \cite{tetko_application_2004-1, xing_novel_2002}. However, a significant gap between logP-based estimation and experimental solubility still remains \cite{delaney_esol:_2004}. Large efforts have been devoted to reduce this gap by adding more input information to the set of descriptors, with a concomitant increase in the complexity of the regressions employed \cite{wang_recent_2011}. 

Two examples of property-based methods, the GSE approach and Delaney's extended GSE (EGSE) approach, rely on two and three fitted parameters, respectively. In \cite{delaney_esol:_2004}, the prediction performances of GSE and EGSE were shown to be $r^2(\mathrm{GSE}) = 0.67$ and $r^2(\mathrm{EGSE}) = 0.69$ for a dataset of 1305 compounds compiled by the authors, which highlights the gap between prediction and experiment for such methodologies.

Structure-based methods predict solubility directly from molecular structural information, which can be implemented by various types of descriptors \cite{hall_electrotopological_1995, rogers_extended-connectivity_2010, durant_reoptimization_2002, duvenaud_convolutional_2015}. Generally, binary fingerprints offer a good trade-off between simplicity and predictive power \cite{zhou_scores_2008, Lind_etal_2003, cheng_binary_2011}. We recently developed the concept of {\sl neural fingerprints} which are structure-based and application-specific with input descriptors generated for arbitrary size and shape based on a molecular graph \cite{duvenaud_convolutional_2015}.

Zhou et al. predicted molecular solubility using a binary circular fingerprint descriptor \cite{zhou_scores_2008}. Although they demonstrated a prediction performance of $r^2 = 0.83$, the authors had to carefully select the training data set in order to achieve that value of $r^2$. Huuskonen showed that a prediction performance of $r^2 = 0.92$ can be achieved by using non-binary descriptors consisting of 53 parameters, including 39 atom-type electro-topological state (E-state) indices \cite{huuskonen_estimation_2000}. However, non-binary descriptors significantly increase computational cost in both the training and validation stages, especially when feature selection is encountered during the regression process \cite{Christoph_Steinbeck_2006, efron_least_2004}. A different binary fingerprint approach has been investigated by Lind and Maltseva, in which support vector regression employing the Tanimotto similarity kernel is applied in order to overcome the limit of the multiple linear regression method \cite{Lind_etal_2003}. 

The group-based methods integrate contributions of all associated functional groups multiplied by the number of each functional group in a compound: $C_0 + \sum_{i=1}^{N} C_i G_i$ where $G_i$ is the number of times the $i$th group appears in the compound, $C_0$ is a constant bias parameter, and $C_i$ is the contribution of the $i$th group \cite{klopman_estimation_1992}. Hou et al. proposed an atom contribution method, which overcomes the 'missing fragment' problem in pure group contribution methods \cite{hou_adme_2004}. The atom contribution method categorizes atoms together with their surrounding molecular environment. Cheng et al. used functional key descriptors such as MACCS Keys and PC881 instead of directing counting numbers of each functional group. This approach simplifies descriptor values to be binary form but 'missing fragment' and requiring a large training data set are still unavoidable for the cases of small and large number of the keys, respectively. Moreover, Cheng et al. apply them for solubility classification task with a much lower solubility requirement, 10 $\mu$g/mL, than the threshold values necessory for aqueous flow battery applications.

% Please add the following required packages to your document preamble:
% \usepackage{multirow}
\begin{table}
\centering
\begin{tabular}{c|c} 
\hline
Category & Methods \\
\hline
\multirow{3}{4em}{Machine Learning} & Property-based method \cite{jain_estimation_2001, ran_estimation_2002, delaney_esol:_2004, Wang_etal_2009}\\  
& Structure-based method \cite{huuskonen_estimation_2000, Lind_etal_2003, zhou_scores_2008}\\  
& Group-based method \cite{klopman_estimation_1992, kuhne_group_1995, cheng_binary_2011}\\ 
\hline
\multicolumn{2}{c}{Quantum Calculation \cite{huuskonen_estimation_2000, mcdonagh_uniting_2014, skyner_review_2015, er_computational_2015}}\\
\hline
\end{tabular}
\caption{A categorization of solubility estimation methods. First, machine learning and quantum calculation methods are depicted. The machine learning methods include the property-based, structure-based and group-based method.}
\label{table:sol_chart}
\end{table}

The ability to carry out solubility predictions that account for pH-dependence is critical to discovering molecules for aqueous flow batteries. In addition to mandating very high solubility, the pH required to operate an organic flow battery system varies depending on the required redox potential values and other experimental considerations. For instance, negative electrolytes of 9,10-anthraquinone-2,7-disulphonic acid (AQDS) in \cite{huskinson_metal-free_2014} and 2,6-dihydroxyanthraquinone (DHAQ) \cite{lin_alkaline_2015} require 1 molar solubility at pH 0 and pH 14, respectively. While prediction methods for intrinsic solubility have been widely discussed, methods to predict pH-dependent solubility have remained less explored \cite{ledwidge_effects_1998, bergstrom_accuracy_2004, hansen_prediction_2006, bhal_rule_2007, skyner_review_2015, wang_silico_2015}. In theory, the Henderson-Hasselbach relationship can be used to predict pH-dependent solubility based on the intrinsic solubility of a molecule \cite{hansen_prediction_2006}. However, the limitations of current pKa prediction accuracies as well as the salt plateau phenomena of ionic solubility encourage the use of a data-driven approaches. This requires significantly more experimental training data (solubility as a function of pH) than intrinsic solubility prediction \cite{bergstrom_accuracy_2004, wang_silico_2015}. Moreover, the intrinsic solubility of extremely strong acids with a negative pKa value has not been well investigated in the literature. 

In high-throughput molecular screening, the development of an accurate and cost-effective property estimation method is a key factor for successfully finding new candidate molecules \cite{pyzer-knapp_what_2015,duvenaud_convolutional_2015,kearnes_scissors:_2014}. In this work, we develop a fast and accurate property estimation method for high-throughput molecular discovery. We named the proposed approach a \emph{multiple descriptor multiple kernel (MultiDK)} method. The method relies on combining an ensemble of different descriptors, including fingerprints, functional keys, as well as other molecular physicochemical properties. We also apply different kernels for different types of descriptors to overcome intrinsic irregularities between a fingerprint and a property \cite{Lind_etal_2003}. Both intrinsic and pH-dependent solubility estimations are supported by the MultiDK approach.

\section{Methods}
\subsection{Datasets and Tools}
We tested the performance of MultiDK on four datasets. The four datasets include 1676 molecules from \cite{wang_development_2007}, 496 molecules from \cite{Willighagen_etal_2006}, 1140 molecules from \cite{delaney_esol:_2004} and 3310 molecules from \cite{Wang_etal_2009}. The 1676 molecule dataset includes most of the 1297 molecules in \cite{huuskonen_estimation_2000}. The tests were performed using 20-fold cross-validation. In this work, we use Python packages including Pandas \cite{mckinney-proc-scipy-2010}, Scikit-learn \cite{scikit-learn}, Tensorflow \cite{tensorflow2015-whitepaper} and Seaborn \cite{michael_waskom_2016_45133} for data manipulation, machine-learning, and visualization tools. 

\subsection{MultiDK method} 
In this paper, we compare the prediction performance of the MultiDK method against single descriptor (SD) and multiple descriptor (MD) methods. The SD method uses only one type of a descriptor, such as a Morgan fingerprint, MACCS keys or a specific molecular physicochemical property. Morgan fingerprints represent an atom and path structure of a molecule using a binary hashing procedure. MACCS keys represent functional group information. For molecular properties, we include molecular weight, Labute's approximate surface area (LASA), or the logarithm partition coefficient (logP). The MD and MultiDK methods include more than one descriptor. Both the Morgan fingerprint and the MACCS keys are binary descriptors while the physicochemical molecular property is a non-binary, real-valued descriptor. 

The MultiDK approach predicts the target molecular property as follows:
\begin{equation}
y = \sum_{i=1,\ldots,L} w^\mathrm{B}_i k_B( \mathbf{x}^\mathrm{B}, \mathbf{x}^\mathrm{B}_{i}) + \mathbf{w}^\mathrm{NB} \mathbf{x}^\mathrm{NB} + w_0
\end{equation}
where $\mathbf{x}^{B}$ and $\mathbf{x}^{NB}$ are binary and non-binary descriptor vectors, respectively. $\mathbf{x}^{B}_{i}$ is a binary descriptor vector for the $i$th training molecule, $\mathbf{w}^{B}$ and $\mathbf{w}^{NB}$ are weight vectors corresponding to $\mathbf{x}^{B}$ and $\mathbf{x}^{NB}$ , respectively, $L$ is the number of a training molecules, and $k_B(\cdot)$ is a binary kernel function.

Rather than using a single kernel or linear regression, MultiDK utilizes multiple kernels such as a nonlinear binary kernel for binary descriptors and linear processing for non-binary descriptors separately. To optimize a kernel function \cite{gonen_multiple_2011,bach_multiple_2004,lanckriet_learning_2004}, multiple combinatorial kernels have been used in various applications including biomedical data \cite{yu_l2-norm_2010} and YouTube video data \cite{chen_event_2013,xu_soft_2013}. Here, we use a multiple kernel approach to apply appropriate kernels for different features instead of training the kernel. The binary kernel function of $k_B(\cdot)$ contributes by exploiting a non-linear relationship between the molecular structure and property. The non-linear relationships arise primarily because each bit indicates the presence or absence of a pattern rather than a quantitative value. MultiDK uses all training molecules as support vector molecules for kernel processing similar to support vector machines. We use the Tanimoto kernel which has been used in a wide range of machine learning applications, such as exploiting binary feature information to recognize white images on a black background \cite{Pekalska_etal_2001} as well as a kernel for support vector and Gaussian progress regression in molecular property prediction \cite{Lind_etal_2003,pyzer-knapp_bayesian_2016}. 

In the MultiDK approach, ensemble learning is employed based on multiple combinational descriptors according to the principle of the 'wisdom of the crowds' \cite{kew_greedy_2015}. The set of descriptors in MultiDK includes the Morgan circular fingerprints \cite{rogers_extended-connectivity_2010}, MACCS Keys \cite{durant_reoptimization_2002} fingerprints and three non-binary molecular properties. The three types of descriptors represent structure hash (atom, path) and structure pattern (key, functional group) and target related molecular properties. We find that this ensemble combination is effective to predict molecular properties because both atom and subgroup representations are employed in the set of descriptors together with the related molecular properties. Moreover, we use different kernels for binary and non-binary descriptors. Particularly, a binary similarity kernel is applied to the binary descriptor and a linear kernel for the non-binary descriptor.       

We evaluate the methods with training and cross-validation phases. In the training phase, we optimize the regression parameters using Ridge regularization. The descriptor consists of 4096 binary bits of the Morgan circular fingerprint with radius 6, 117 binary bits of the MACCS Keys and a few non-binary scalar descriptors. We generate all descriptors using the RDKit tool \cite{rdkit2015} except for the partition coefficient, which we obtain from  \emph{Cxcalc} from the Chemaxon Marvin suite \cite{cxcalc_2009}. Before linear regression, we pre-process the 4213 binary bits with the binary similarity kernel by calculating the Tanimoto similarity between an input vector and the set of training vectors. We pass the non-binary descriptors directly to the linear regression stage without pre-processing. Then, the binary kernel output values and the direct non-binary output value are entered into the Ridge linear regression stage. We employ the Ridge regression routine in the scikit-learn Python package \cite{sklearn_api}.
The regularization process eventually produces the best regression coefficients and an intercept corresponding to the maximum $R^2$ performance. In the cross-validation phase, a combination vector of the binary kernel outputs and a direct descriptor of a test molecule is multiplied by the coefficients obtained in the training phase.

\subsection{MultiDK for estimating intrinsic solubility, logS}
We use MultiDK for solubility prediction as follows:
\begin{equation}
\log S = \sum_{i=1,\ldots,L} w^\mathrm{CK}_i k_B( \mathbf{x}^\mathrm{CK}, \mathbf{x}^\mathrm{CK}_{i}) + (\mathbf{w}^\mathrm{WSP} \cdot \mathbf{x}^\mathrm{WSP}) + w_0
\end{equation}
where the subindices C, K, W, S, and P represent the Morgan circular fingerprint, the MACCS keys, the molecular weight, Labute's approximate surface area (LASA) (Labute 2000) and the logarithm partition coefficient (logP), respectively. $L$ is the number of a training molecules, $k_B( \cdot)$ is a binary kernel function, $\mathbf{x}^{MCMK} = [\mathbf{x}^{MC}, \mathbf{x}^{MK}]$ is a concatenated binary vector for an input molecule, $\mathbf{x}^{a}_{i}$ is a concatenated binary vector of the $i$th supporting molecule, and $x^\mathrm{MW}$ is molecular weight (MW). Both $w^\mathrm{MCMK}_i$ and $w^\mathrm{MW}$ are regression coefficients and $w_0$ is the regression intercept. The values of $\mathbf{x}^\mathrm{MC}$, $\mathbf{x}^\mathrm{MK}$ and $x^\mathrm{MW}$ are generated according to the SMILES string of a molecule.

\subsection{MultiDK for estimating pH dependent solubility, logS(pH)}
In order to predict pH-dependent solubility, we extend the MultiDK method as follows:  
\begin{equation} \label{eq:logS_pH}
\log S(\mathrm{pH}) = \log S + \log P - \log D( \mathrm{pH})
\end{equation}
where $\log P$ and $\log D(\mathrm{pH})$ are the $n$-octanol-to-water partition coefficient and the pH-dependent distribution coefficient, respectively. Since the two coefficients can be approximated as $\log P = \log S_\mathrm{Oct} - \log S$ and $\log D(\mathrm{pH}) = \log S_\mathrm{Oct} - \log S(\mathrm{pH})$ \cite{jain_estimation_2001, ran_estimation_2002}, we are able to extend MultiDK as in (\ref{eq:logS_pH}) where $\log S_\mathrm{Oct}$ is solubility in octanol. The octanol solubility is intrinsic and therefore determined regardless of existence of ionizable groups \cite{sijm_aqueous_1999}. We evaluate both $\log P$ and $\log D( \mathrm{pH})$ using the \emph{cxcalc} plugin in the Chemaxon Marvin suite \cite{chemicalizer_2015}. 

\section{Results and Discussion}
\subsection{Cross-validation results}

\subsubsection{Performance of MultiDK for solubility prediction}
\begin{figure}
\begin{center}
\includegraphics[width=0.7\columnwidth]{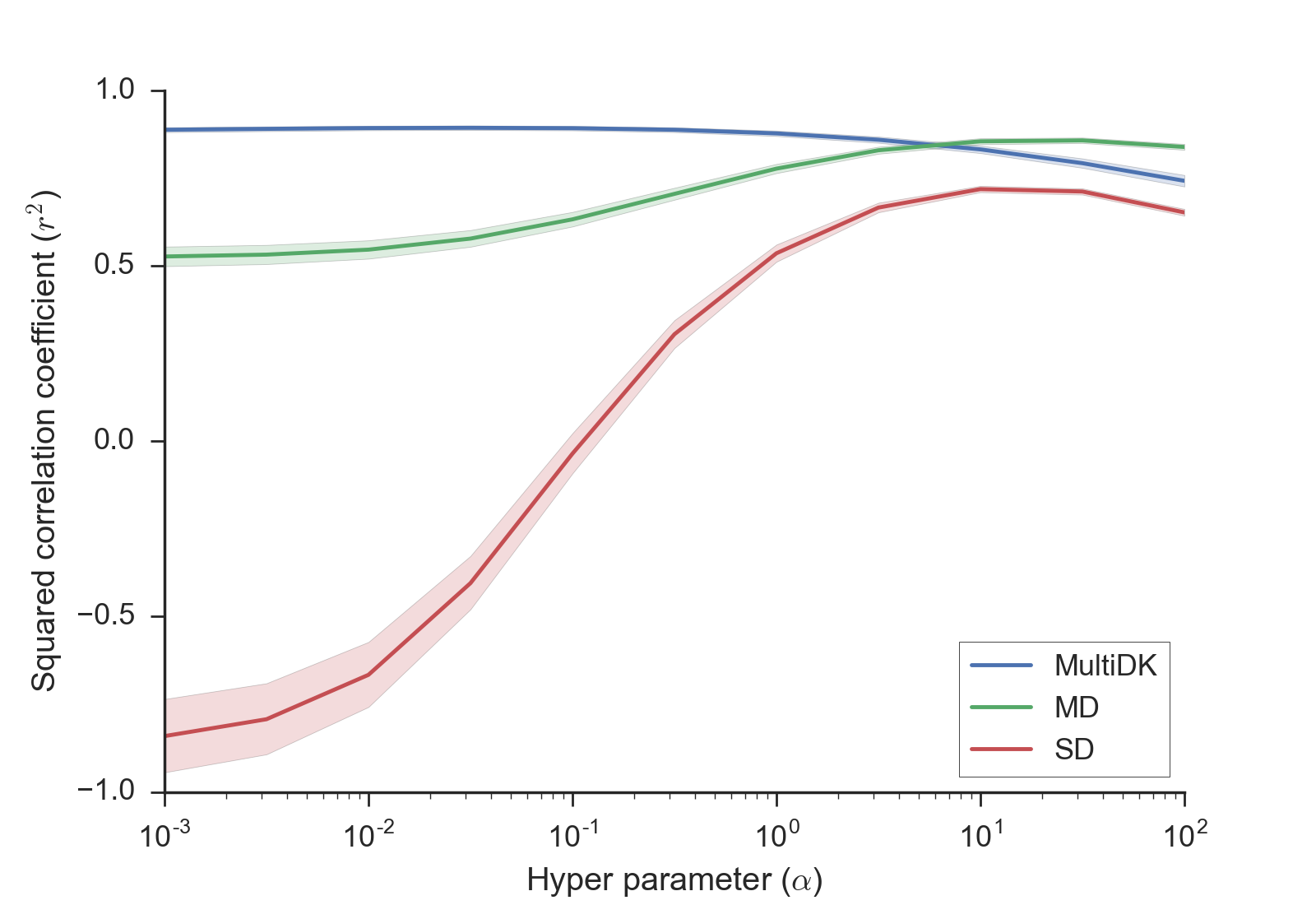}
\caption{\label{fig:logS_alpha} Solubility prediction as a function of the Ridge regression hyperparameter $\alpha$ for the SD, MD and MDMK cases. For each $\alpha$ in $10^{-3}$ to $10^2$, a 20-fold cross-validation was applied.%
}
\end{center}
\end{figure}

We use $r^2$ distribution of 20-fold cross validation as a metric of prediction performance. The $r^2$ distribution is obtained by 20 time repetition of both training and testing until 20 subsets of data are all used for validation. Figure \ref{fig:logS_alpha} shows the $r^2$ distribution obtained with each of the methods tested as a function of the Ridge regression hyper-parameter $\alpha$. Here, we used the 1676 unique molecules in \cite{wang_development_2007}. For efficient comparison, only one non-binary descriptor is considered in this evaluation. Both the MultiDK and the MD methods employ two binary and one non-binary descriptors where the two binary and one non-binary descriptors are Morgan fingerprints (MFP), MACCS Keys (MACCS) and molecular weight (MolW). As shown in the figures, MultiDK and MD significantly outperform SD. Moreover, MultiDK is most robust to changes in the value of $\alpha$. This result reveals that additional group and property information help improve the regression performance.   

\begin{figure}
\begin{center}
\includegraphics[width=0.7\columnwidth]{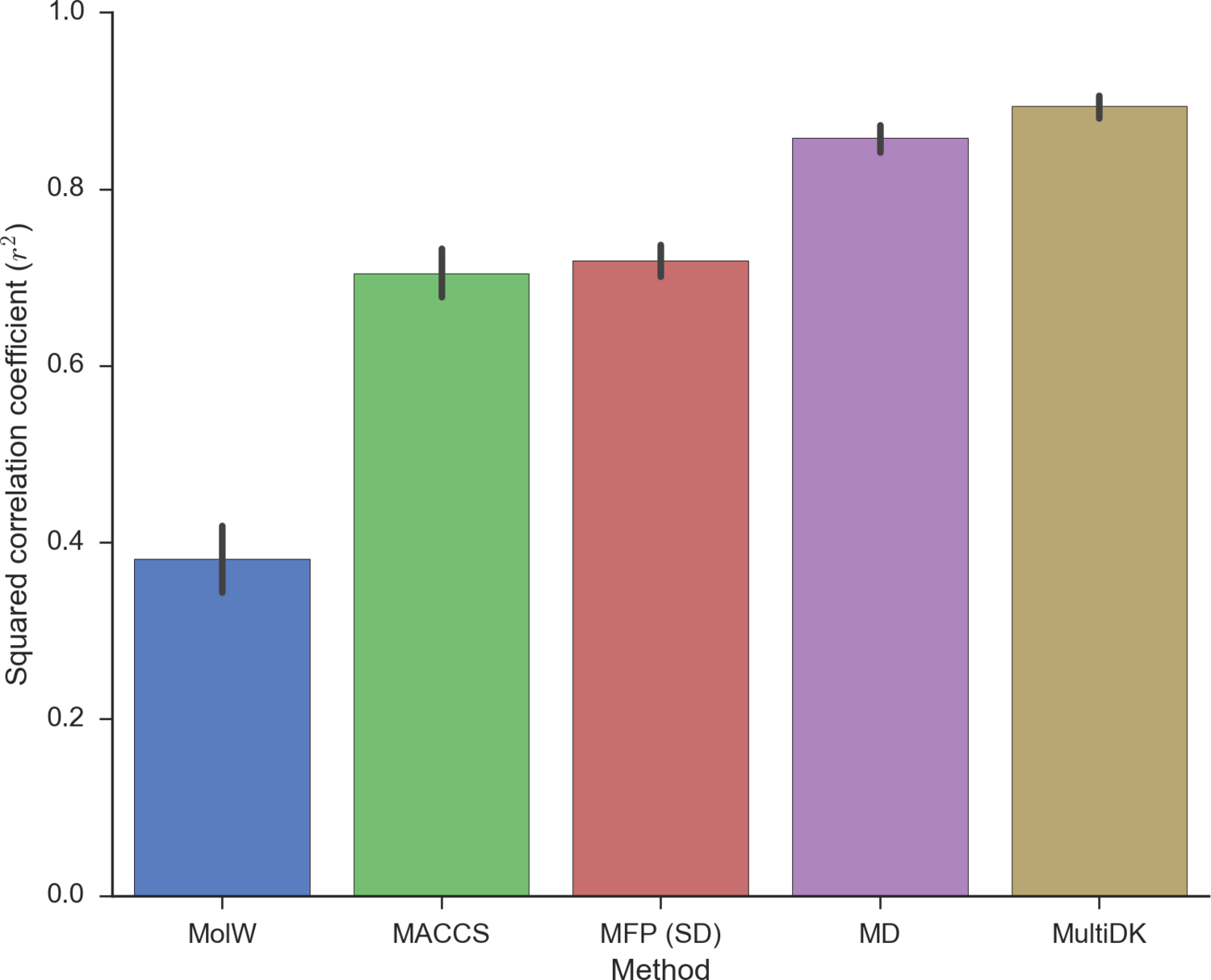}
\caption{\label{fig:MDMK-1708-bar}, the performances of SD, MD and MDMK are compared when the optimal value of $\alpha$ is used. For SD, molecular weights (MolW), MACCS Keys (MACCS) and Morgan fingerprint (MFP or SD) are independently used as a descriptor.%
}
\end{center}
\end{figure}

In Figure \ref{fig:MDMK-1708-bar}, the performances of SD family, MD and MDMK are compared when the optimal value of $\alpha$ is used, where the SD family includes MFP, MACCS and MolW. This bar graph shows a clear difference between the SD family, MD and MultiDK approaches. The best $\alpha$ value are found by a grid search approach which selects $\alpha$ on the basis of regression performance in the range of $10^{-3}$ to $10^{2}$ with 10 logarithmically equally spaced steps. Each regression performance is evaluated using a 20-fold cross-validation with initial data shuffling. SD (MFP), MD and MDMK achieve their best regression coefficient values of $r^2$ $\pm$ std($r^2$) = 0.72 $\pm$ 0.04, 0.86 $\pm$ 0.04 and 0.89 $\pm$ 0.03 at $\alpha=$ 10.0, 31.6 and 0.03, respectively. This result highlights three important points. First, SD with MFD outperforms the other two SDs approaches, SD using MACCS and SD using MolW. It suggests that detailed structural information helps to estimate solubility. MolW is one non-binary value and MACCS and MFB consist of 117 and 4069 binary values, respectively. Second, both MD and MultiMK outperform SD, which emphasizes the necessity of multiple type descriptors for accurately estimating molecular properties. Third, MultiDK can further improve prediction performance in comparison to MD through the use of a binary kernel regression.

\subsubsection{Performance of MultiDK with more descriptors}
\begin{figure}
\begin{center}
\includegraphics[width=0.7\columnwidth]{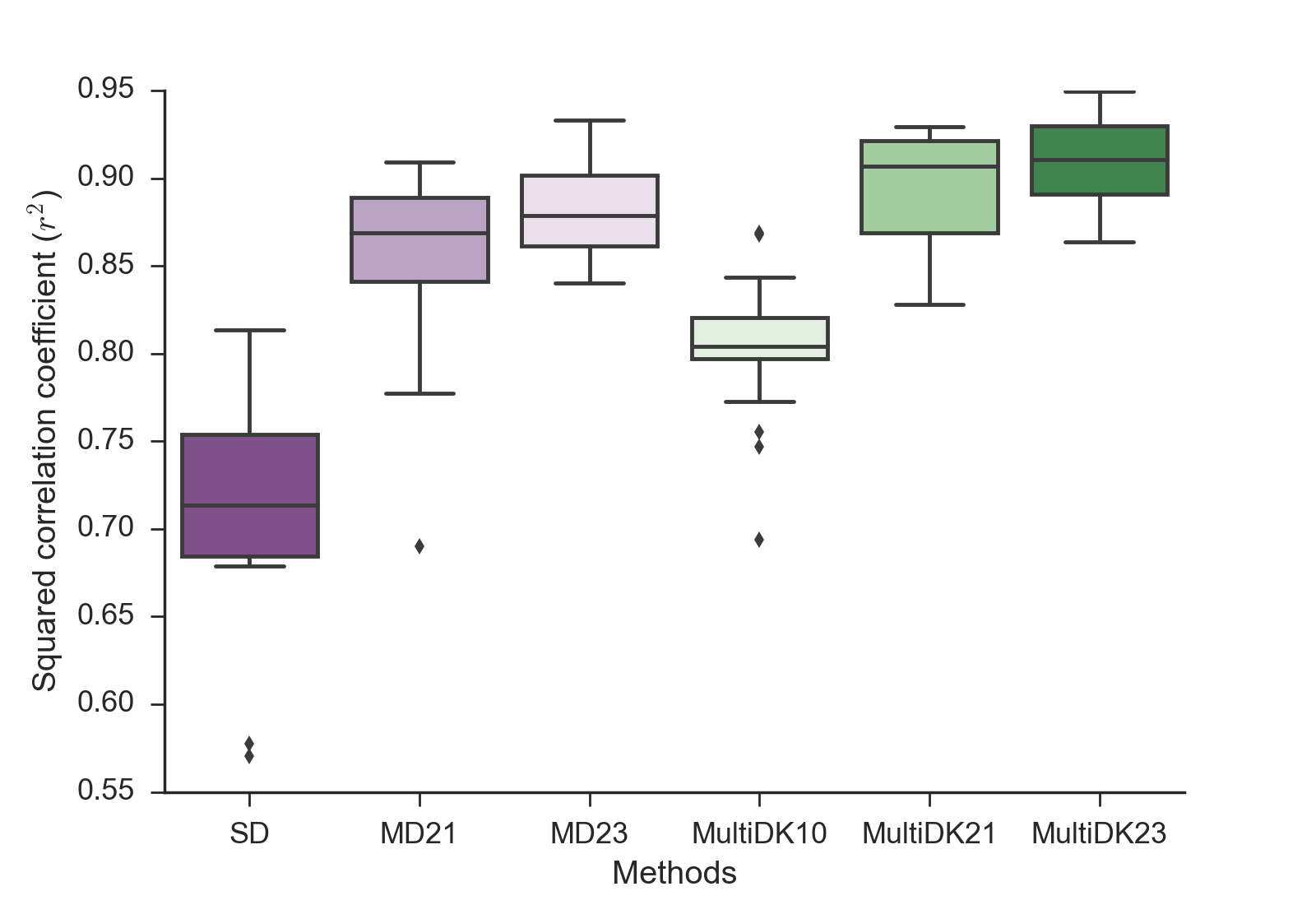}
\caption{\label{fig:Wang1708_box}Prediction performance of different methods with the dataset with 1676 molecules.%
}
\end{center}
\end{figure}

\begin{figure}
\begin{center}
\includegraphics[width=0.7\columnwidth]{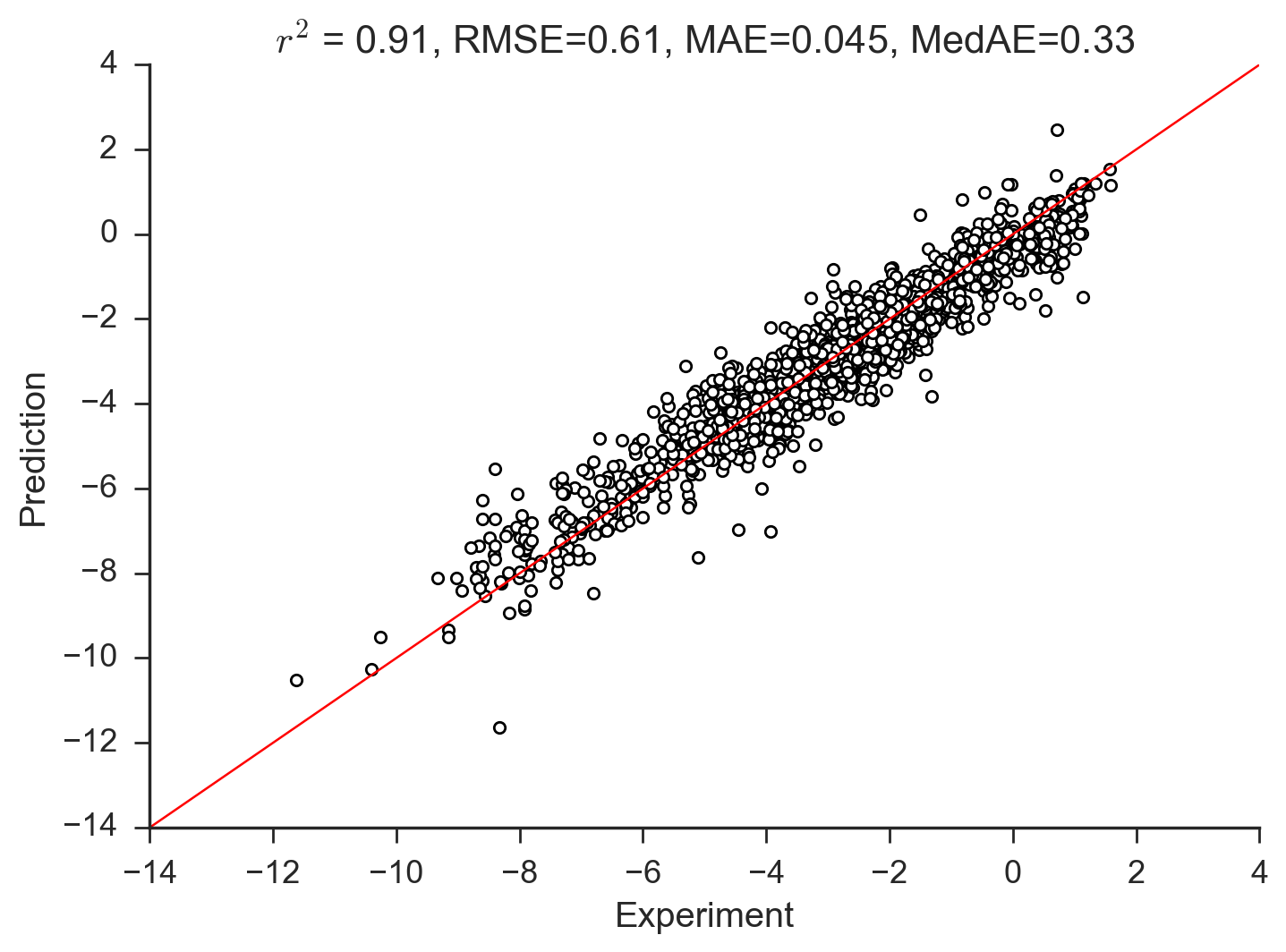}
\caption{\label{fig:MDMK-cv1708} Comparision of the 1676 experimental solubility data and cross-validation results of MultiDK using the optimal value of $\alpha$.%
}
\end{center}
\end{figure}

The $r^2$ distributions of different methods on the 1676 molecules using more descriptors are shown in Figure \ref{fig:Wang1708_box} where the box represents the interquartile range of $r^2$ values, i.e., the difference between the first quartile and the second quartile, and the median of them is drawn inside the box. The numerical values of them are shown in Table \ref{table:Wang1708_bar}. We include two more non-binary descriptors which are Labute's approximate surface area (LASA) \cite{labute_widely_2000} and the logarithm partition coefficient (logP). Paricularly for MultiDK, we include a method with separate binary kernels for each binary descriptor. MD$xy$ and MultiDK$xy$ represent a method which embeds $x$ binary and $y$ non-binary descriptors. Figure \ref{fig:MDMK-cv1708} shows a comparison of the experimental data and the MultiDK results obtained through cross-validation with the best $\alpha$. We obtained the following cross-validation summary statistics: mean($r^2$) = 0.91, std($r^2$) = 0.027, root mean squared error (RMSE) = 0.61, mean absolute error (MSE) = 0.45, median absolute error (MSE) = 0.33.

\begin{table} 
\centering
\begin{tabular}{l|ccc}
\hline 
Method & Best $\alpha$ & E[$r^2$] & std($r^2$) \\
\hline 
SD          & 1E+1 & 0.72 & 0.06 \\
MD21        & 3E+1 & 0.86 & 0.05 \\
MD23        & 3E+1 & 0.88 & 0.03 \\
MultiDK10   & 1E-3 & 0.80 & 0.04 \\
MultiDK21   & 3E-2 & 0.89 & 0.04 \\
MultiDK23   & 1E-1 & 0.91 & 0.03 \\
\hline
\end{tabular}
\caption{20-fold cross-validation performances of the 1676 molecules} 
\label{table:Wang1708_bar}
\end{table}

\subsubsection{Performance of MultiDK for other datasets}
\begin{figure}
\begin{center}
\includegraphics[width=0.7\columnwidth]{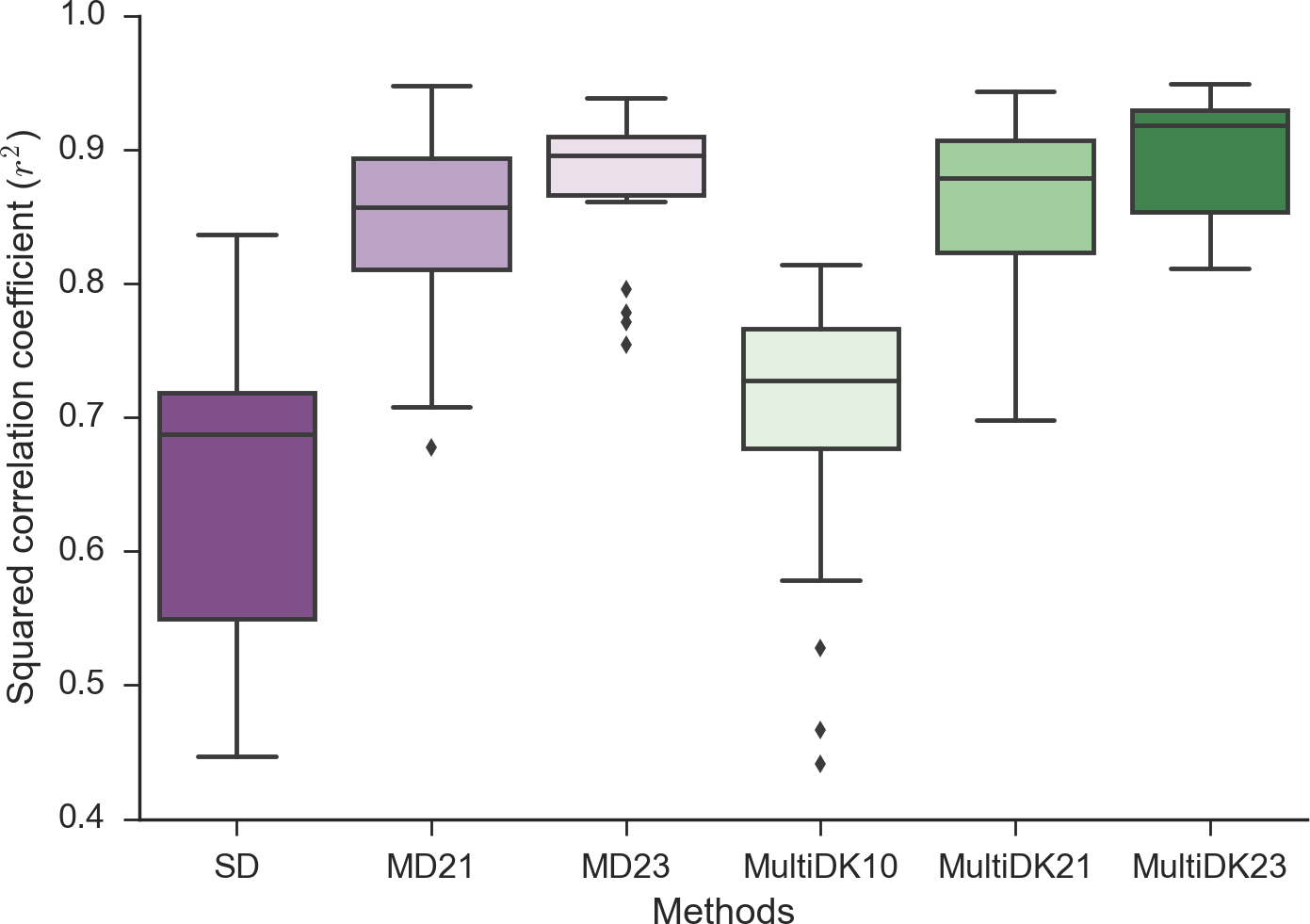}
\caption{\label{fig:WS496_box}Prediction performance of different methods with the dataset with 496 molecules.%
}
\end{center}
\end{figure}

\begin{figure}
\begin{center}
\includegraphics[width=0.7\columnwidth]{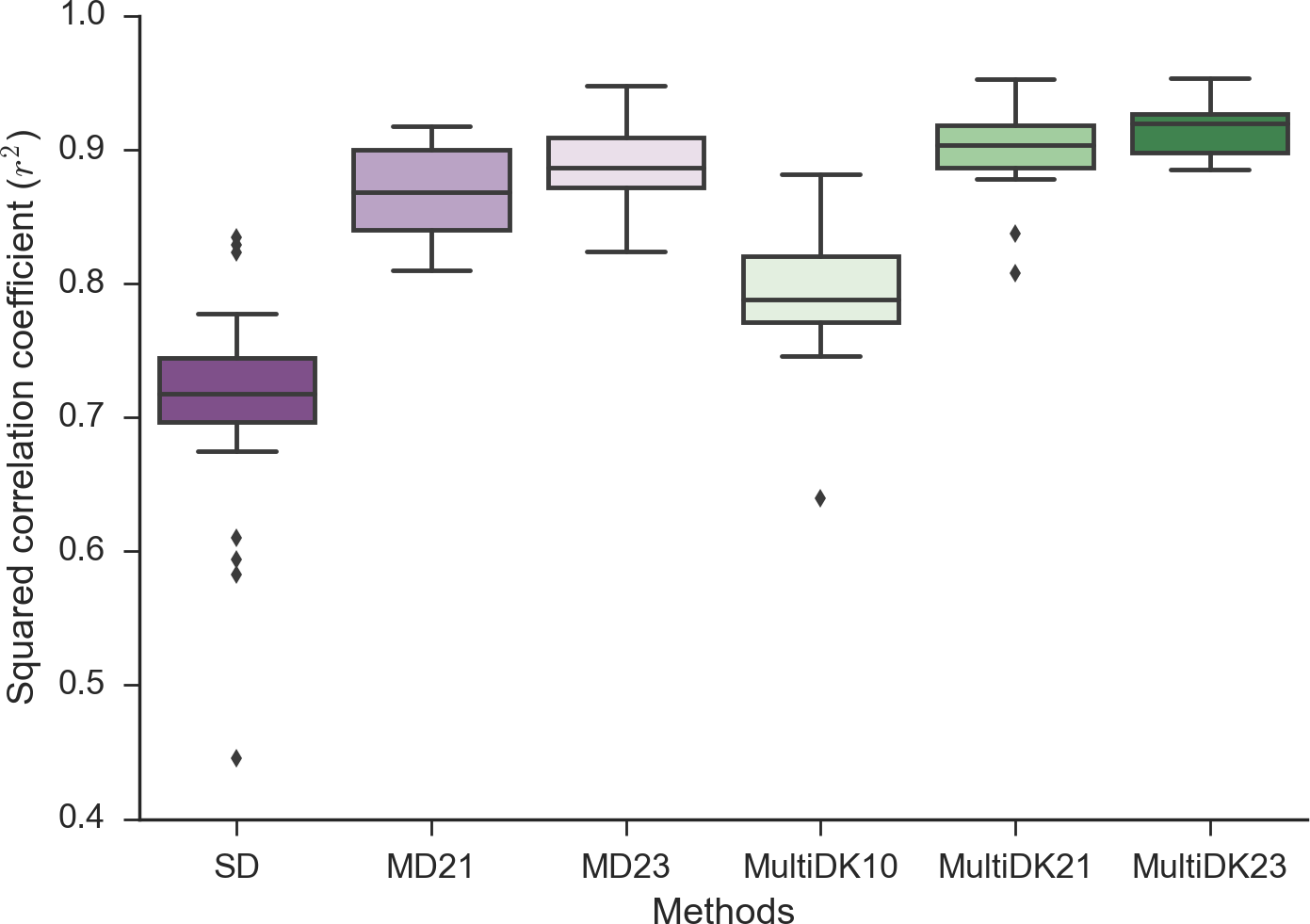}
\caption{\label{fig:D1144_box}Prediction performance of different methods with the dataset with 1140 molecules.%
}
\end{center}
\end{figure}

\begin{figure}
\begin{center}
\includegraphics[width=0.7\columnwidth]{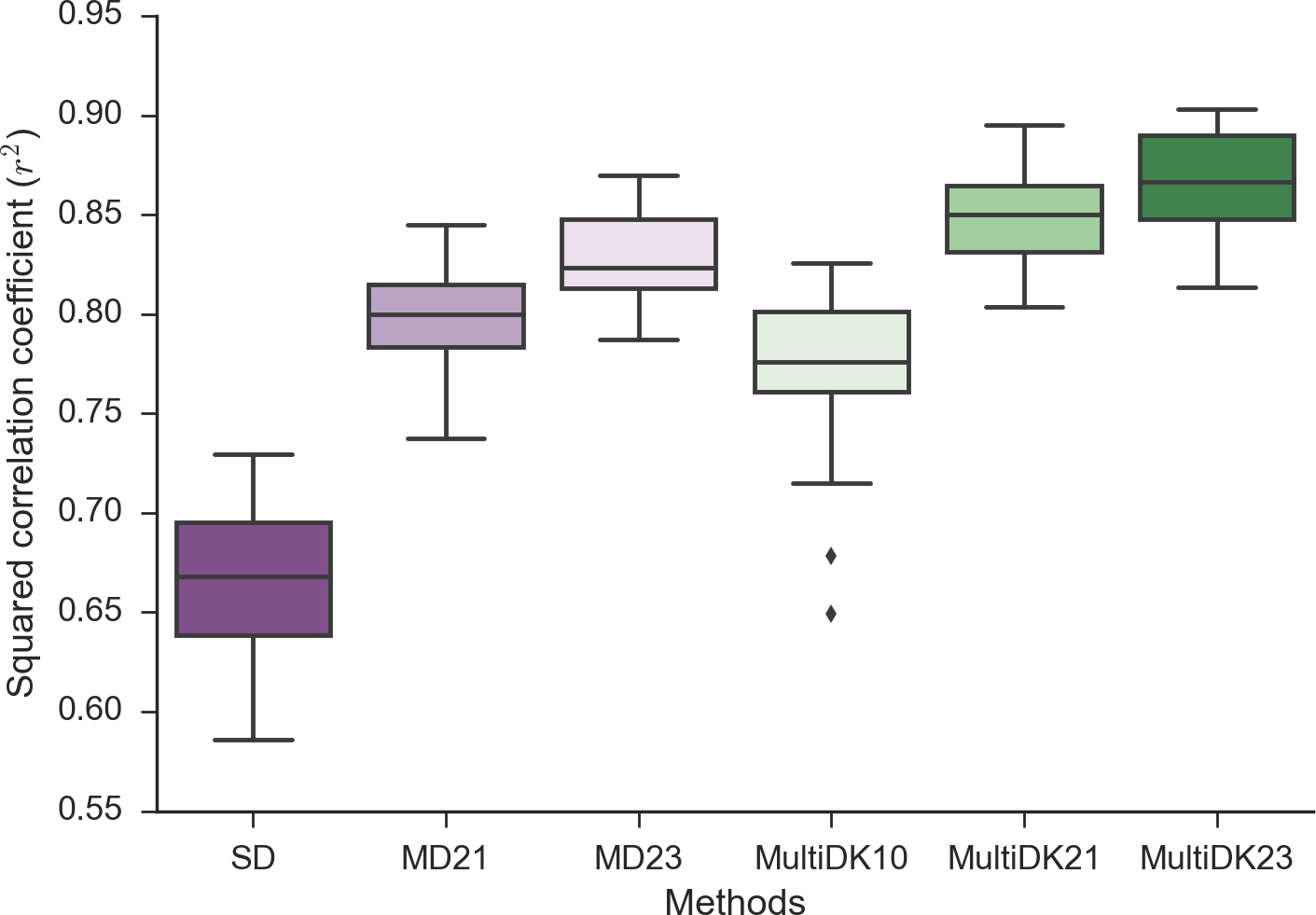}
\caption{\label{fig:Wang3705_box}Prediction performance of different methods with the dataset with 3310 molecules.%
}
\end{center}
\end{figure}

The three more datasets of 496 molecules \cite{Willighagen_etal_2006}, 1140 molecules \cite{delaney_esol:_2004} and 3310 molecules \cite{Wang_etal_2009} are considered in order to verify the proposed MultiDK method as shown in Figures \ref{fig:WS496_box}, \ref{fig:D1144_box}, and \ref{fig:Wang3705_box}, respectively. The average values and standard deviation of $r^2$ obtained across multiple cross validation iterations are illustrated in Table \ref{table:3datasets_bar}. From the figures and the table, we confirm that the performance of MultiDK are better than MD for all new three data sets when the same input descriptors are used. Moreover, SD with only MFP is shown to be the worst among all cases, which is equivalent to the previous 1676 molecule case. 

\begin{table}
\centering
\begin{tabular}{l|ccc|ccc|ccc}
\hline 
\multirow{2}{4em}{Method} 
            & \multicolumn{3}{c|}{496 molecules}  & \multicolumn{3}{c}{1140 molecules}   & \multicolumn{3}{|c}{3310 molecules}  \\
        & Best $\alpha$ & E[$r^2$] & std($r^2$) & Best $\alpha$ & E[$r^2$] & std($r^2$) & Best $\alpha$ & E[$r^2$] & std($r^2$) \\
\hline
SD        & 1E+1     & 0.65      & 0.12   & 1E+1     & 0.71      & 0.09   & 3E+1     & 0.66      & 0.05   \\
MD21      & 1E+1     & 0.84      & 0.07   & 1E+1     & 0.87      & 0.04   & 3E+1     & 0.79      & 0.06   \\
MD23      & 1E+1     & 0.88      & 0.06   & 1E+1     & 0.89      & 0.03   & 3E+1     & 0.83      & 0.02   \\
MultiDK10 & 3E-3     & 0.70      & 0.11   & 3E-3     & 0.79      & 0.05   & 3E-2     & 0.77      & 0.05   \\
MultiDK21 & 7E-2     & 0.86      & 0.06   & 3E-2     & 0.90      & 0.04   & 1E-1     & 0.85      & 0.03   \\
MultiDK23 & 7E-2     & 0.89      & 0.05   & 3E-2     & 0.92      & 0.02   & 1E-1     & 0.87      & 0.04  \\
\hline
\end{tabular}
\caption{Performances of solubility prediction for different datasets}
\label{table:3datasets_bar}
\end{table}

\subsection{Application to the prediction of quinone electrolytes}
\subsubsection{Intrinsic solubility prediction of quinone molecules}
\begin{figure}
\begin{center}
\includegraphics[width=1\columnwidth]{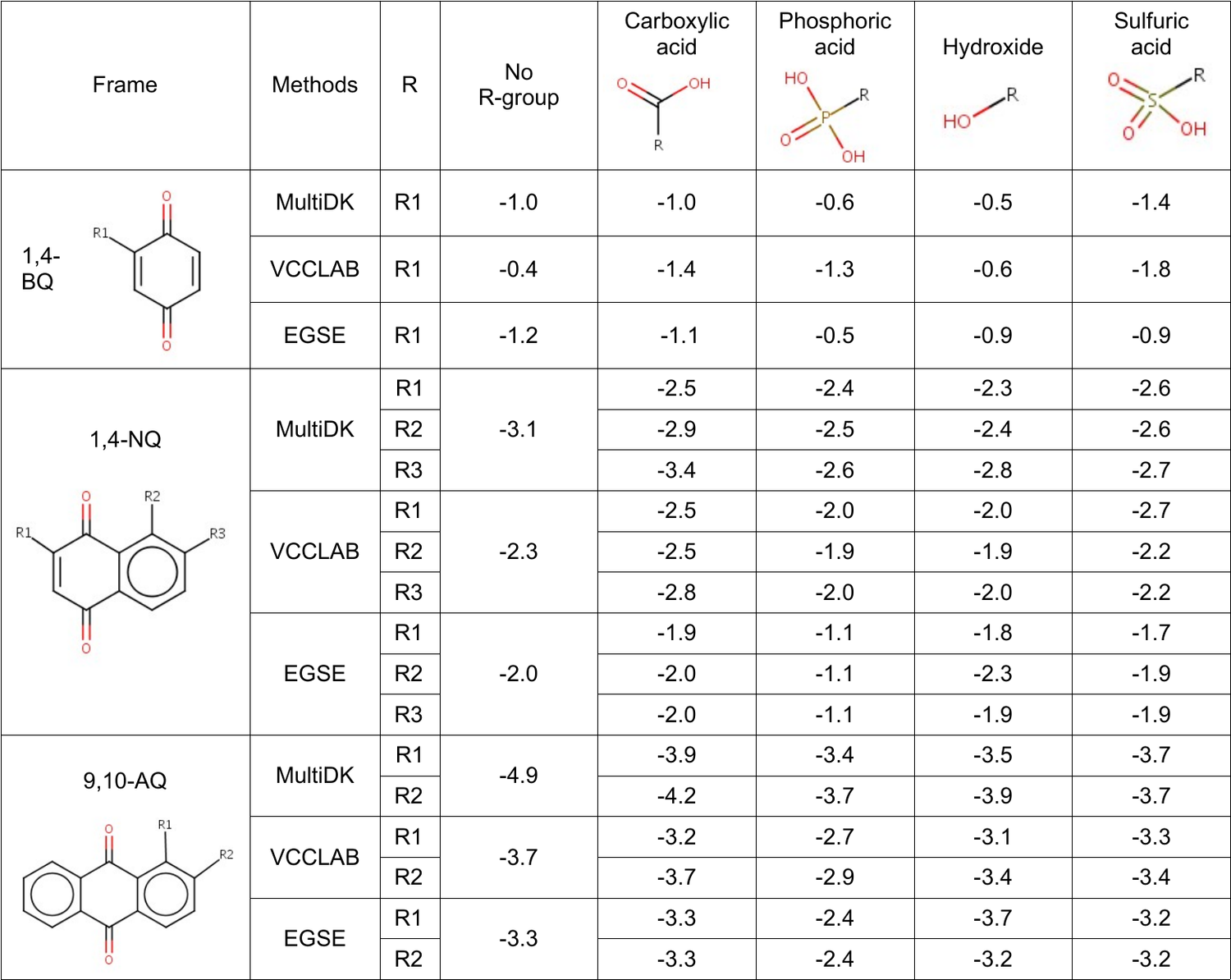}
\caption{\label{fig:bnaq-r} Predicted solubility of 27 quinone molecules by three different methods, i.e., MultiDK, VCCLAB and EGSE, where Benzoquinone (BQ), naphthoquinone (NQ) and anthraquinone (AQ), with available unique positions of R-group attachment.%
}
\end{center}
\end{figure}

Next, we apply the MultiMK method to predict the solubility of a set of quinone molecules, which are useful electrolytes for organic aqueous flow batteries. The intrinsic solubility is defined as the solubility of a molecule in its neutral form. Three types of quinone families, benzoquinones (BQ), naphthoquinones (NQ) and anthraquinones (AQ) as shown in Figure \ref{fig:bnaq-r}, were considered.  

\begin{figure}
\begin{center}
\includegraphics[width=0.7\columnwidth]{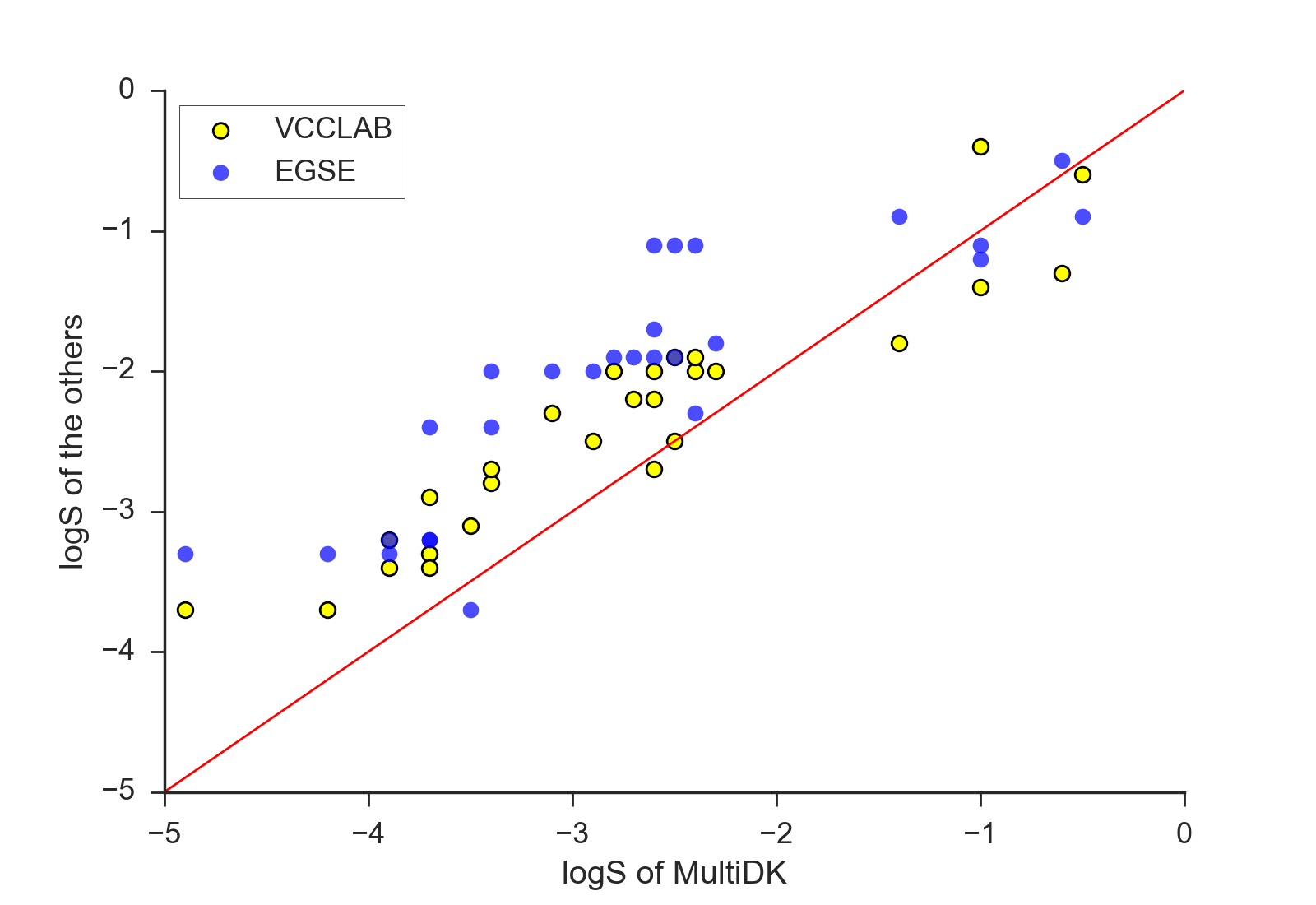}
\caption{\label{fig:bike_vcclab_egse} Three sets of predicted solubility values for 27 quinones compared against each other. Solubility values were predicted using the MultiDK, VCCLAB and EGSE methods. The three methods show that the predicted intrinsic solubility values of the 27 quinones are lower than 0 log molar, regardless of the attached functional group. 0 log molar is the general solubility requirement of electrolytes for inexpensive organic aqueous flow battery applications.%
}
\end{center}
\end{figure}

We tested molecules belonging to the BQ, NQ and AQ with no or one substituent R-group, where the number of the total test molecules are 27 consisting 5 BQ, 13 NQ and 9 AQ family molecules. The intrinsic solubility was predicted using three different methods: MultiDK, VCCLAB and EGSE. VCCLAB is an on-line solubility estimation tool (http://www.vcclab.org/lab/alogps/) and EGSE estimates the intrinsic solubility as:
\begin{equation} \label{eq:egse}
\log S = 0.16 - 0.63 C \log P - 0.0062 \mathrm{MW} + 0.066 \mathrm{RB} - 0.74 \mathrm{AP}
\end{equation}
where MP is the melting point, MW is the molecular weight, RB is a rotational bond ,and AP is an aromatic portion of the molecule. VCCLAB estimates solubility by training 1291 molecules using an artificial neural network \cite{tetko_estimation_2001}. As shown in Figure \ref{fig:bike_vcclab_egse}, regardless of the molecule types or the attached R-groups, all three methods predict the intrinsic solubility (logS) of the molecules to be below zero log-molar. Thus, all molecules have intrinsic solubility less than the solubility target of the aqueous flow battery.

\subsubsection{pH-dependent solubility for single R-group quinones} 
\begin{figure}
\begin{center}
\includegraphics[width=0.7\columnwidth]{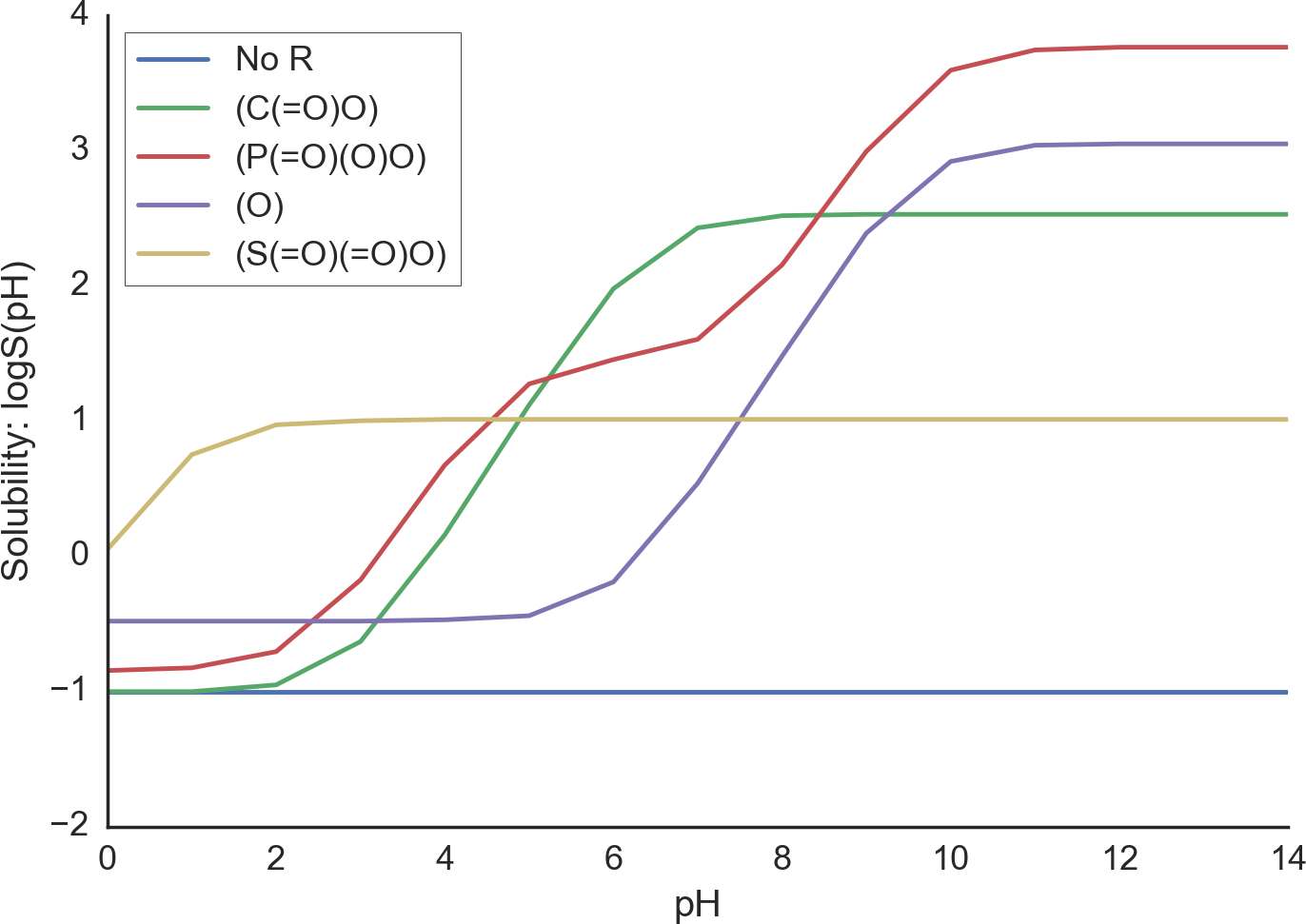}
\caption{\label{fig:BQ_PlogS_pH} Predicted pH-dependent solubility of benzoquinones (BQ) with different functional groups. The legend describes R groups enumerated with BQ. Depending on pH, the solubility values of the quinones with a R-group significantly vary.%
}
\end{center}
\end{figure}

\begin{figure}
\begin{center}
\includegraphics[width=0.7\columnwidth]{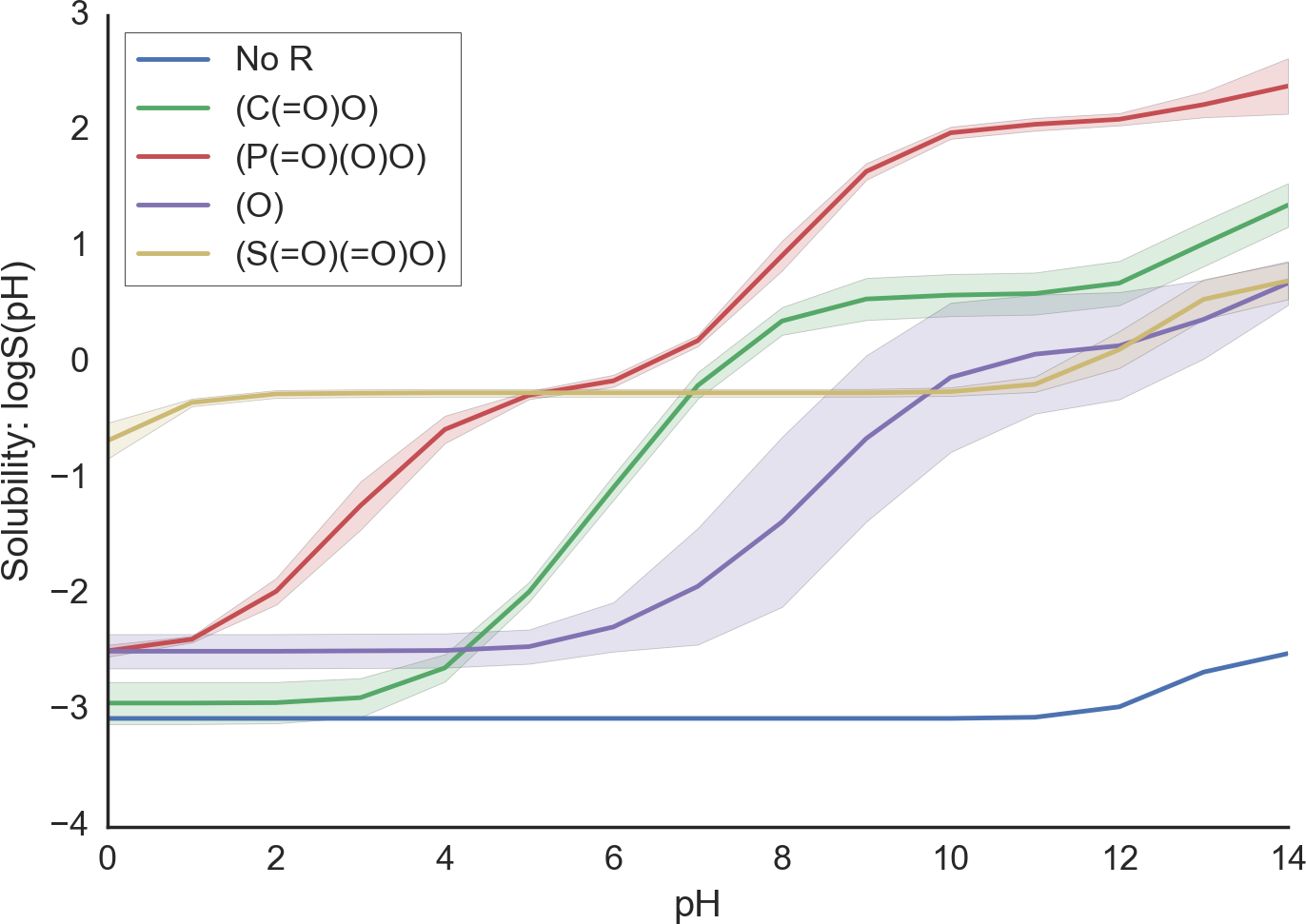}
\caption{\label{fig:NQ_PlogS_pH} Predicted pH-dependent solubility of naphthoquinones (NQ) with different functional group substituents. Three unique positions are available to attach functional groups in NQ.%
}
\end{center}
\end{figure}

\begin{figure}
\begin{center}
\includegraphics[width=0.7\columnwidth]{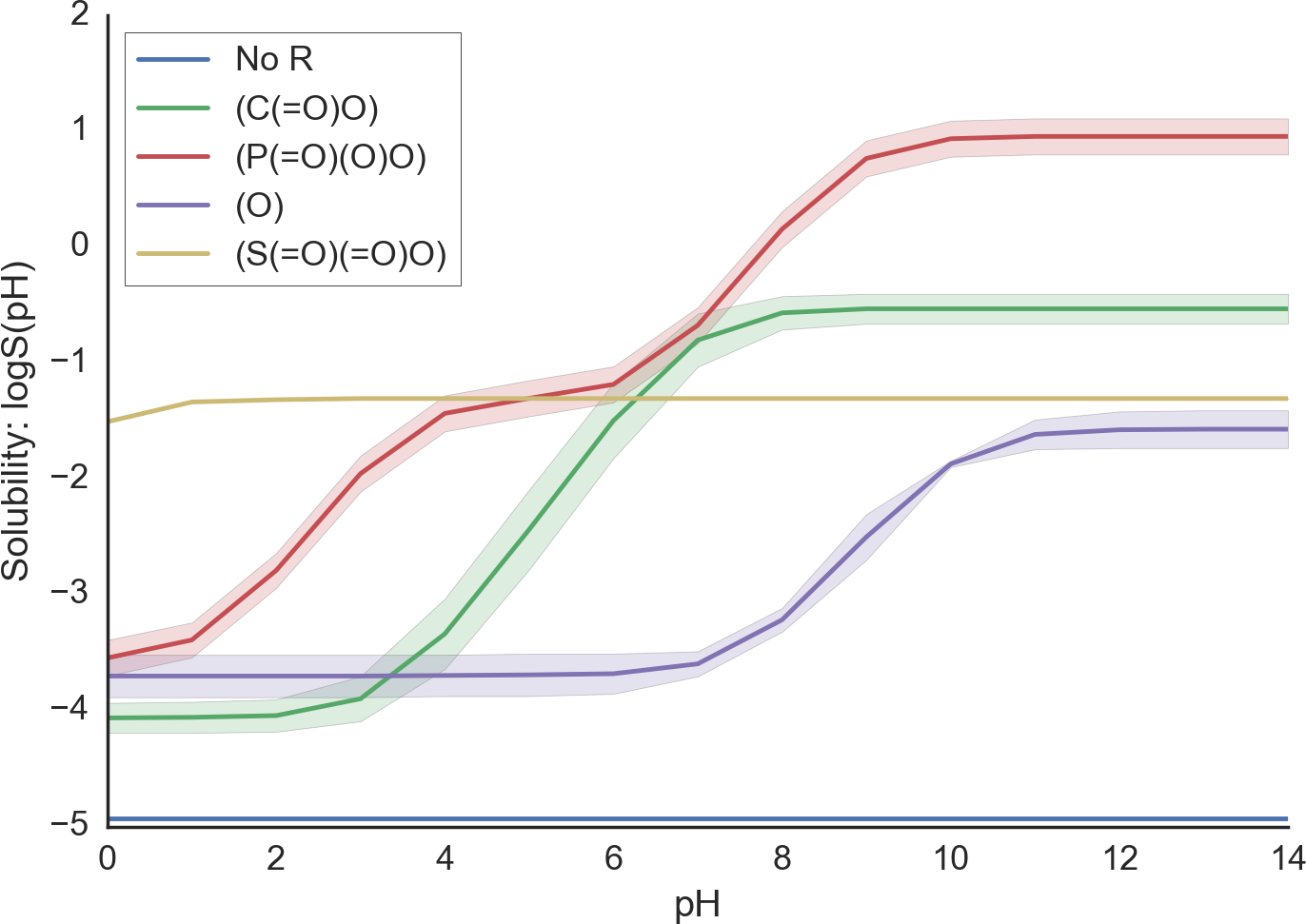}
\caption{\label{fig:AQ_PlogS_pH} Predicted pH-dependent solubility of anthraquinone (AQ) with different functional group substituents. Two unique positions are available to attach functional groups in AQ.%
}
\end{center}
\end{figure}

In Figure \ref{fig:BQ_PlogS_pH}, \ref{fig:NQ_PlogS_pH} and \ref{fig:AQ_PlogS_pH}, we show pH-dependent solubility predicted by the extended MultiDK method. We applied the extended method to the three types of quinone family molecules. Figure \ref{fig:BQ_PlogS_pH} shows the predicted pH-dependent solubility for five BQ molecules which are BQ with a sulfonic acid (SO$_3$H), phosphori acid (PO$_3$H), carboxylic acid (COOH) and hydroxide (OH) or no R group. The BQ with a sulfonic acid, phosphoric acid, carboxylic acid are shown to be the best soluble molecules at at pH=0, 7, and 14, respectively. 

Figure \ref{fig:NQ_PlogS_pH} shows predicted pH-dependent solubility of 13 NQ molecules which are NQ with one of the same four R-group to the BQ case or no R group. Figure \ref{fig:AQ_PlogS_pH} shows the predicted pH-dependent solubility of 9 AQ molecules which are AQ with one of the same four R-group to the BQ and NQ cases or no R group. Both the NQ and AQ with a sulfonic acid and phosphori acid are shown to be the best soluble molecules at at pH=0 and 14, respectively, while both the NQ and AQ with hydroxide and no R-group are less soluble than the other molecules at pH=7.

\begin{figure}
\begin{center}
\includegraphics[width=0.7\columnwidth]{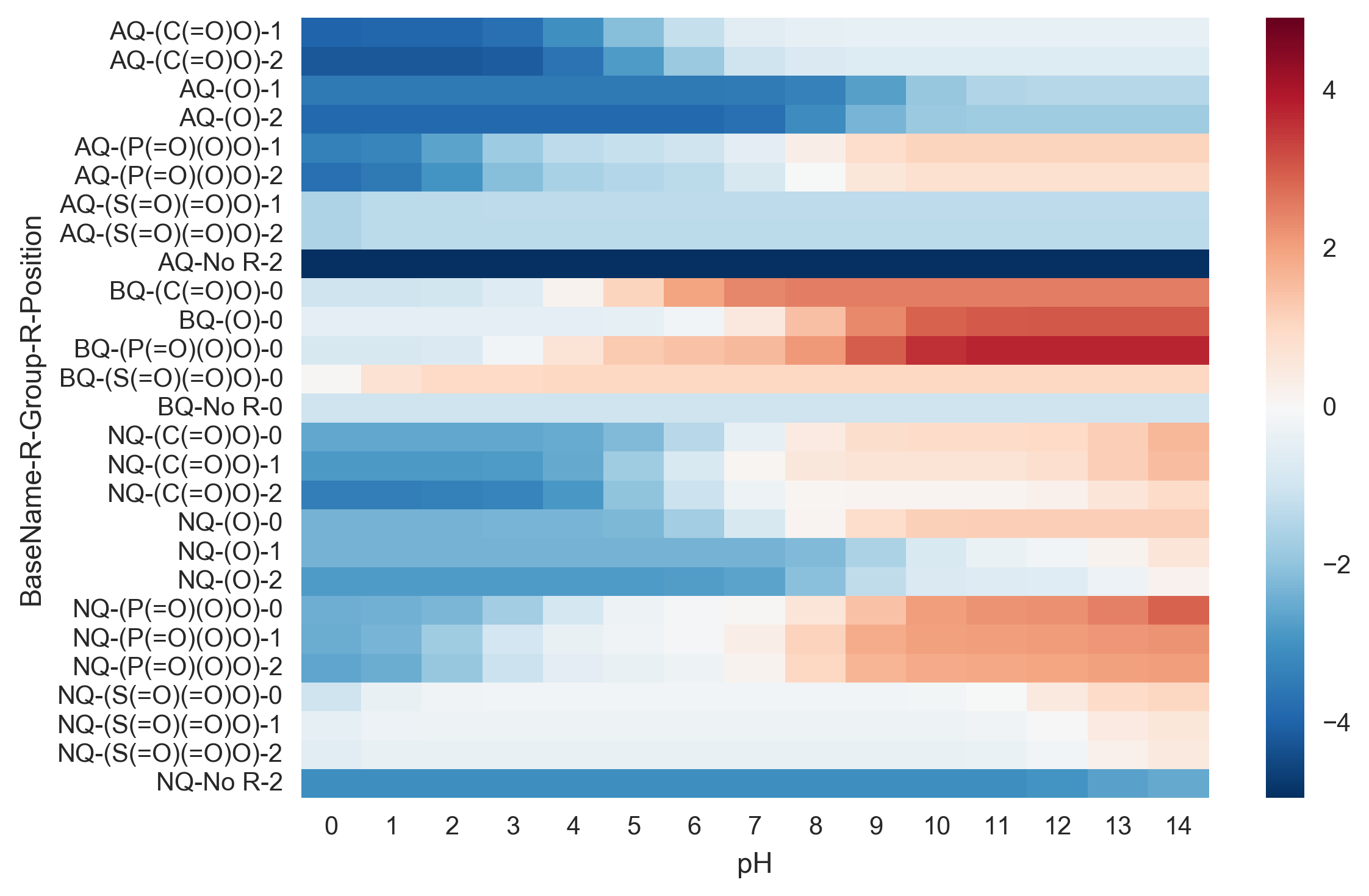}
\caption{\label{fig:BNA_PlogS_pH} Heatmap of predicted pH-dependent solubility of all three quinone families with different functional group substituents.%
}
\end{center}
\end{figure}

\subsubsection{pH-dependent solubility of multiple R-group anthraquinones}
\begin{figure}
\begin{center}
\includegraphics[width=0.84\columnwidth]{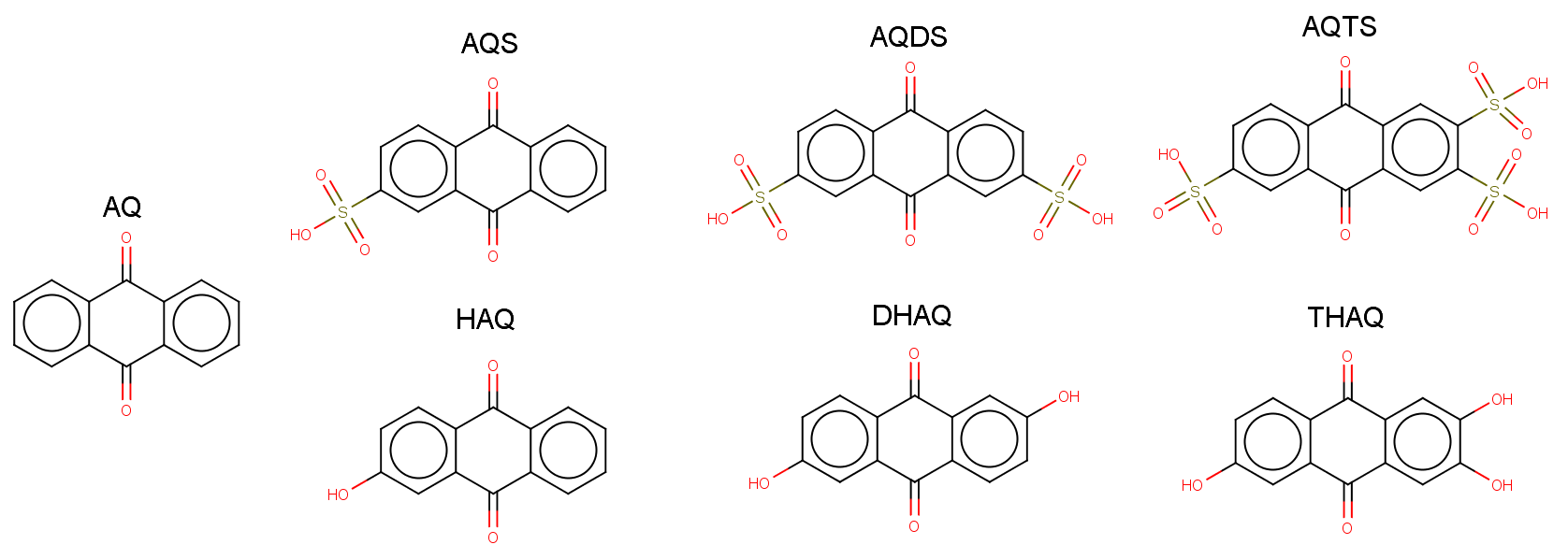}
\caption{\label{fig:AQ_AQxS_xHAQ}  Anthraquinone and anthraquinone with either zero, mono-, di- and tetra-sulfonic acid or hydroxyl groups. Anthraquinone (AQ), anthraquinonesulfonic acid (AQS), anthraquinone-disulfonic acid (AQDS),  anthraquinone-tetrasulfonic acid (AQTS), hydroxyl-anthraquinone (HAQ), dihydroxyl-anthraquinone (DHAQ) and tetrahydroxyl-anthraquinone (THAQ) are illustrated.
}
\end{center}
\end{figure}

We predict the pH-dependent solubility of quinone molecules with multiple R-groups. Particularly, anthraquinone with multiple sulfonic acid groups and multiple hydroxyl groups are considered. Figure \ref{fig:AQ_AQxS_xHAQ} shows structures of anthraquinones with zero, one, two and three sulfonic acid or hydroxyl groups. Quinone molecules with attached sulfonic acid group are particularly interesting since they display high solubilities and desirable redox potential values. In particular, 9,10-anthraquinone-2,7-disulphonic acid was chosen as a negative electrolyte \cite{huuskonen_estimation_2000} and 1,2-dihydrobenzoquinone- 3,5-disulfonic acid was selected as a positive electrolyte \cite{yang_inexpensive_2014} for the acid quinoe flow batteries. The alkaline quinone flow battery embodies 2,6-dihydroxy-9,10-anthraquinone (2,6-DHAQ) as a negative electrolyte, and the experiment solubility of 2,6-DHAQ is reported as more than 0.6 M in 1 M KOH \cite{lin_alkaline_2015}.

\begin{figure}
\begin{center}
\includegraphics[width=0.84\columnwidth]{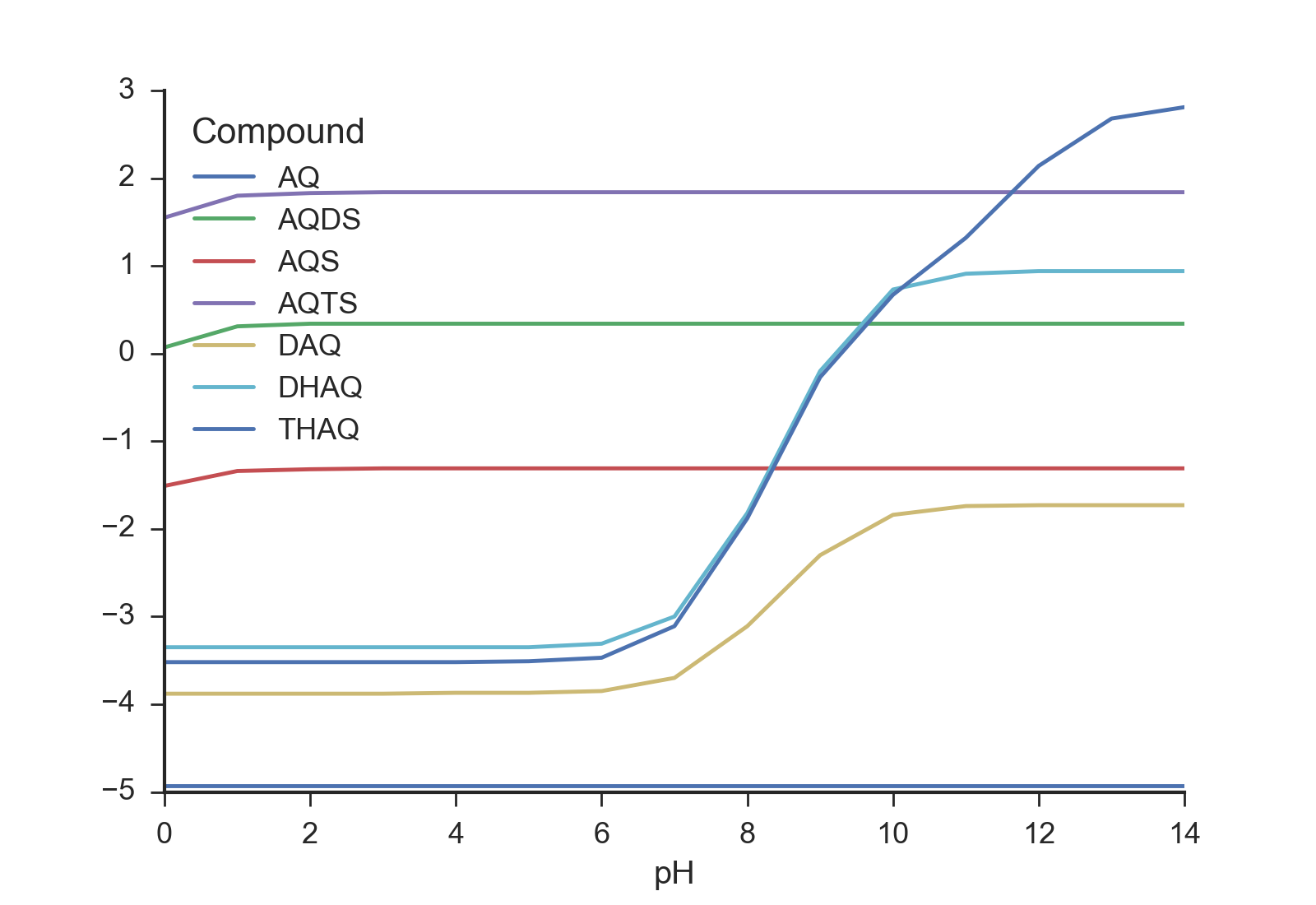}
\caption{\label{fig:AQ4} Predicted intrinsic and pH dependent solubility of seven anthraquinone family molecules with sulfonic or hydroxyl groups. Although their intrinsic solubility is predicted to have similar values, their pH-dependent solubility values are significantly varied depending on how many and which functional groups are attached.%
}
\end{center}
\end{figure}

Figure \ref{fig:AQ4} show that anthraquinone with no such R-groups is far insoluble in any pH condition while Table \ref{table:AQxS_AQxH} picks solubility at pH 0, 7, 14 and includes prediction results by Chemaxon Cxcalc with logS plug-in as well as the extended MultiDK method. The MultiDK prediction shows that more sulfonic acid groups, more soluble, such as $P \log S_\mathrm{pH}$(AQTS) $>$  $P \log S_\mathrm{pH}$(AQDS) $>$ $P \log S_\mathrm{pH}$(AQS) $\gg$ $P \log S_\mathrm{pH}$(AQ), in all pH condition including the acid case and more hydroxyl groups, more soluble, such as $P \log S_\mathrm{pH}$(THAQ) $>$  $P \log S_\mathrm{pH}$(DHAQ) $>$ $P \log S_\mathrm{pH}$(HAQ) $\gg$ $P \log S_\mathrm{pH}$(AQ), in alkali condition. Therefore, it is noteworthy that an efficient prediction method should clearly differentiate between the solubility of an enumerated molecule according to the number of ionic functional groups in every pH points. The MultiDK with pH-dependent solubility estimation can be used as a more practical tool than the intrinsic solubility prediction method especially for the application of dicoverying organic flow battery electrodes.

\begin{table}
\centering
\caption{pH-dependent solubility of AQ with multiple R-groups where sulfonic acid and hydroxyl groups are considered. The pH-dependent solubility of them are estimated by MultDK and Chemaxon Cxcalc.}
\label{table:AQxS_AQxH}
\begin{tabular}{c|ccc|ccc}
    \hline 
       & \multicolumn{3}{c|}{MultiDK} & \multicolumn{3}{c}{Cxcalc}\\
    pH       &  0  & 7  & 14 & 0  & 7  & 14\\
    \hline 
    AQ       & -4.9  & -4.9 & -4.9 & -4.5 & -4.5 & -4.5\\
    \hline 
    AQS      & -1.5  & -1.3 & -1.3 & -1.6 &    0 & 0\\
    AQDS     &  0.1  & 0.3  & 0.3  & 0 & 0 & 0 \\
    AQTS     &  1.6  & 1.8  & 1.8  & 0 & 0 & 0\\
    \hline 
    HAQ      & -3.9  & -3.7 & -1.7 & -4.1 & -3.9 & 0\\
    DHAQ     & -3.4  & -3.0 & 0.9  & -3.7 & -3.3 & 0\\
    THAQ     & -3.5  & -3.1 & 2.8  & -3.3 & -2.9 & 0\\
    \hline
\end{tabular}
\end{table}

\section{Conclusion}
Organic aqueous flow battery systems require highly soluble electrolytes, which are two- to five-fold more soluble than pharmaceutical drugs. In order to search molecules with such a tight solubility requirement, high-throughput screening is a compelling approach especially when it is combined with an efficient solubility prediction method. Moreover, the investigation of pH-dependent solubility is essential to discovery highly soluble molecules which include an ionizable fragment such as the sulfonic acid (-SO$_3$H), the carboxylic acid (-COOH), the hydroxyl (-OH) and the dihydrogen phosphite (-PO$_3$H$_2$). We have developed a multiple descriptor multiple kernel (MultiDK) approach as an efficient property prediction method. As the ensemble descriptor consists of structure hash and fragment keys fingerprints as well as one or a few property specific descriptors such as molecular weight only or additionally Labute's approximate surface area, and a partition coefficient, it has shown that MultiDK is capable of fast, accurate and universal solubility prediction. By the extension of MultiDK, the pH-dependent solubility of various quinones even with strong acidic or alkaline functional groups was investigated at each pH point where the quinones are the strong candidates of electrolytes for organic aqueous flow batteries.

\section*{Acknowledgement}
This work was funded by the U.S. DOE ARPA-E award DE-AR0000348. We thank Roy G. Gordon and Michael J. Aziz for helpful discussions. The support of Changwon Suh and Rafael G{\'o}mez-Bombarel was useful in this work. 

\bibliography{bibliography/biblio.bib%
}

\newpage
\section{Supplementary Information}
\subsection{MultiDK vs. SVR and DNN}
The performance of support vector regression (SVR) and deep neural network (DNN) are tested for solubility estimation. The same descriptors to the cases of MultiDK23 are used for them. We evaluate SVR and DNN using the Scikit-learn and the Tensorflow packages in Python, respectively. 

\begin{figure}
\begin{center}
\includegraphics[width=0.7\columnwidth]{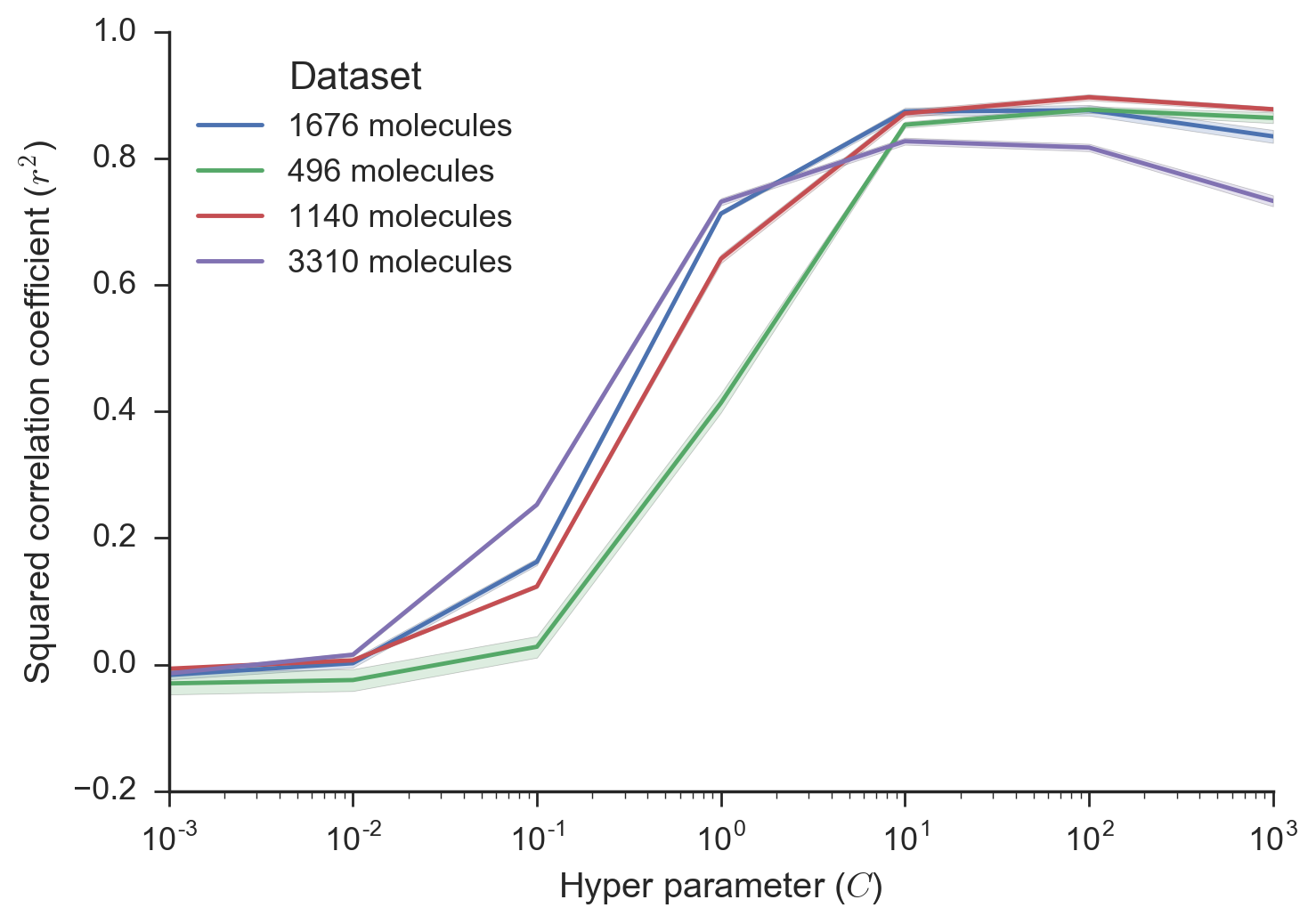}
\caption{\label{fig:svr_r2} The $r^2$ distributions of SVR with respect to the hyper parameter of $C$.%
}
\end{center}
\end{figure}

For SVR, we choose the kernel as radial basis function (RBF), which is given by
\begin{equation}
k_\mathrm{RBF}(\mathbf{x}, \mathbf{y}) = e^{\gamma | \mathbf{x}-\mathbf{y}|^2}
\end{equation}
Penalty hyper parameter $C$ is searched for seven logarithmically equal spaced points from 1E-3 to 1E+3, while the other hyper parameter $\epsilon$ specifying the epsilon-tube and $\gamma$ are adjusted by the default values provided in the Scikit-learn package: $\epsilon = 0.1$ and $\gamma = 1/\mathrm{\#features}$. It is noteworthy that SVR requires a float point kernel computation for all descriptors regardless of a descriptor type while MultiDK computes binary kernel operation which obviously significantly faster than a float point computation. Figure \ref{fig:svr_r2} and Table \ref{table:svr_r2} show that MultiDK outperforms RBF-SVR in all data set cases. Particularly, the average $r^2$ values of MultiDK and RBF-SVR are 0.87 and 0.83, respectively.   

\begin{table} 
\centering
\begin{tabular}{c|ccc|ccc}
\hline 
\multirow{2}{4em}{Method} 
            & \multicolumn{3}{c|}{SVR}  & \multicolumn{3}{c}{MultiDK} \\
        & Best $C$ & E[$r^2$] & std($r^2$) & Best $\alpha$ & E[$r^2$] & std($r^2$) \\
\hline 
1676 molecules & 1E+2 & 0.88 & 0.02 & 1E-1 & 0.91 & 0.03\\
496 molecules  & 1E+2 & 0.87 & 0.04 & 7E-2 & 0.89 & 0.05\\
1140 molecules & 1E+2 & 0.90 & 0.01 & 3E-2 & 0.92 & 0.02\\
3310 molecules & 1E+1 & 0.83 & 0.01 & 1E-1 & 0.87 & 0.04\\
\hline
\end{tabular}
\caption{Average and std of the best $r^2$ values of SVR for each data set} 
\label{table:svr_r2}
\end{table}

For DNN, we evaluate the largest data set which includes the 3310 molecules. Also 20\% of them are used for external testing while the 20\% of the remained molecules are used for internal validation for DNN. We applied a lot of different network architectures manually and eventually find that a three hidden layer DNN with 100, 50, 10 weights for the first, second and third hidden layers shows the best performance among all our test structures. The performance of the best DNN is $r^2$=0.84, RMSE=0.86, MAE=0.60, DAE=0.42 for the test molecules, which is worse than the average $r^2$ of MultDK23 whereas DNN also employs the same descriptors to those of MultDK23, as aforementioned. The DAE represnts median absolute error. 

\begin{figure}
\begin{center}
\includegraphics[width=0.7\columnwidth]{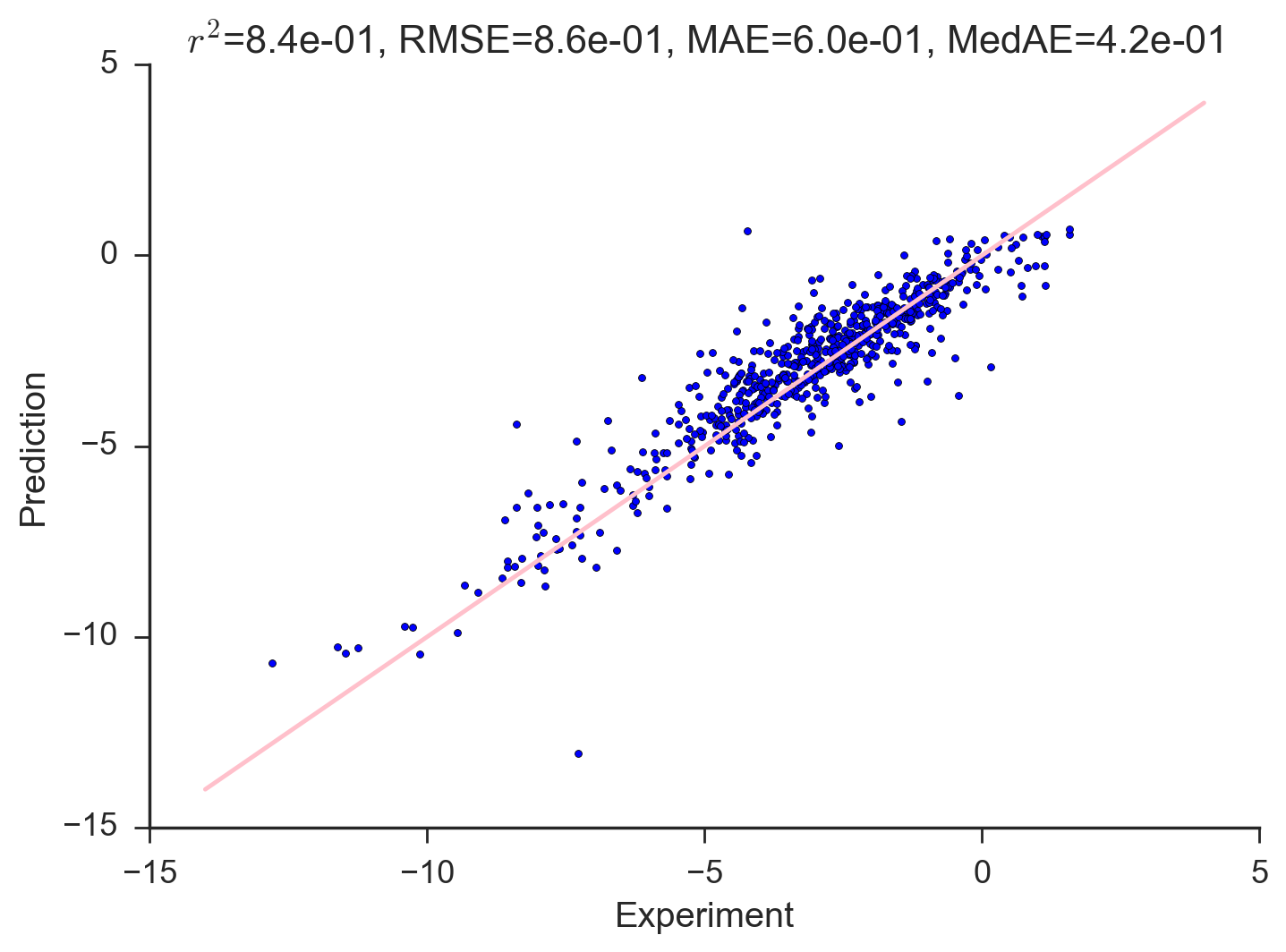}
\caption{\label{fig:DL_test} The experimental and predicted solubility of DNN for the test molecules are compared.
}
\end{center}
\end{figure}

\subsection{Kernels for a binary descriptor}
The Tanimoto similarity has been used as a kernel function to exploit binary feature information such as recognizing white images on a black background. For further understanding, we compare the Tanimimoto similarity kernel with the linear kernel. The linear kernel is given by
\begin{equation}
k_\mathrm{L}( \mathbf{x}, \mathbf{x}^a_i) = \mathbf{x}^T \mathbf{x}^a_i = s
\end{equation}
and the Tanimoto similarity kernel is given by
\begin{equation}
k_\mathrm{T}( \mathbf{x}, \mathbf{x}^a_i) = \frac{ f_{\wedge}(\mathbf{x}, \mathbf{x}^a_i)}{ f_{\vee}(\mathbf{x}, \mathbf{x}^a_i)} =  \frac{s}{s+d} = \frac{1}{1+d/s} 
\end{equation}
where both $s = \mathbf{x}^T \mathbf{x}_i$ and $f_{\wedge}(\mathbf{x}, \mathbf{x}^a_i) = \sum_{j} x_j \wedge x^a_{i,j} = s$ are both the number of common 1's in two vectors, $f_{\vee}(\mathbf{x}, \mathbf{x}^a_i) = \sum_{j} x_j \vee x^a_{i,j}$ is the number of 1's in any two vectors and $d$ is equal to $f_{\vee}(\mathbf{x}, \mathbf{x}^a_i) - s$. The linear kernel of $k_\mathrm{L}( \mathbf{x}, \mathbf{x}^a_i)$ does not rely on $d$, while $k_\mathrm{T}( \mathbf{x}, \mathbf{x}^a_i)$ is inversely proportional to $d$ similar to a characteristic of the radial basis function. Therefore, a kernel regression with $k_\mathrm{T}( \mathbf{x}, \mathbf{x}^a_i)$, the Tanimoto similarity, can offer better performance than the linear kernel regression as shown in the main text, refering to MD versus MultiDK. 

\end{document}